\documentclass[11pt,a4paper]{article}

\usepackage{amssymb}
\usepackage{graphicx}
\usepackage{bm}

\usepackage{mathrsfs}
\usepackage{cite}

\begin{document}

\title{Quantum properties of gauge theories with extended supersymmetry formulated in ${\cal N}=1$ superspace}

\author{S.S.Aleshin\,${}^{a}$, K.V.Stepanyantz\,${}^{bcd}$ $\vphantom{\Big(}$
\medskip\\
${}^a${\small{\em Institute for Information Transmission Problems RAS,}}\\
\medskip
{\small{\em 127051, Moscow, Russia,}}\\
${}^b${\small{\em Moscow State University, Faculty of Physics,}}\\
{\small{\em Department of Theoretical Physics,}}\\
\medskip
{\small{\em 119991, Moscow, Russia,}}\\
${}^c${\small{\em Moscow State University, Faculty of Physics,}}\\
{\small{\em Department of Quantum Theory and High Energy Physics,}}\\
{\small{\em 119991, Moscow, Russia}}\\
\vphantom{1}\vspace*{-2mm}\\
$^d$ {\small{\em Bogoliubov Laboratory of Theoretical Physics, JINR,}}\\
{\small{\em 141980 Dubna, Moscow region, Russia.}}\\
\vphantom{1}\vspace*{-2mm}
\medskip
}

\maketitle

\begin{abstract}
We analyse quantum properties of ${\cal N}=2$ and ${\cal N}=4$ supersymmetric gauge theories formulated in terms of ${\cal N}=1$ superfields and investigate the conditions imposed on a renormalization prescription under which the non-renormalization theorems are valid. For this purpose in these models we calculate the two-loop contributions to the anomalous dimensions of all chiral matter superfields and the three-loop contributions to the $\beta$-functions for an arbitrary ${\cal N}=1$ supersymmetric subtraction scheme supplementing the higher covariant derivative regularization. We demonstrate that, in general, the results do not vanish due to the scheme dependence, which becomes essential in the considered approximations. However, the two-loop anomalous dimensions vanish if a subtraction scheme is compatible with the structure of quantum corrections and does not break the relation between the Yukawa and gauge couplings which follows from ${\cal N}=2$ supersymmetry. Nevertheless, even under these conditions the three-loop contribution to the $\beta$-function does not in general vanishes for ${\cal N}=2$ supersymmetric theories. To obtain the purely one-loop $\beta$-function, one should also chose an NSVZ renormalization prescription. The similar statements for the higher loop contributions are proved in all orders.
\end{abstract}

\section{Introduction}
\hspace*{\parindent}

Non-renormalization theorems essentially improve the ultraviolet behaviour of supersymmetric theories in comparison with the non-supersymmetric case. In particular, in ${\cal N}=1$ supersymmetric theories the superpotential does not receive divergent quantum corrections \cite{Grisaru:1979wc}, so that the renormalizations of masses and Yukawa couplings appear to be related to the renormalization of chiral matter superfields. Moreover, the $\beta$-function of supersymmetric theories is related to the anomalous dimension of the chiral matter superfields by the NSVZ equation \cite{Novikov:1983uc,Jones:1983ip,Novikov:1985rd,Shifman:1986zi}. In theories with extended supersymmetry the cancellation of divergences is much more significant. In particular, in ${\cal N}=2$ supersymmetric gauge theories all contributions to the $\beta$-function beyond the one-loop approximation vanish \cite{Grisaru:1982zh, Howe:1983sr,Buchbinder:1997ib}. Also the anomalous dimensions of all chiral superfields in these theories are equal to 0 \cite{Howe:1983sr,Howe:1983wj,Buchbinder:1997ib}. Therefore, choosing a gauge group and a representation for the hypermultiplet superfields in such a way that the one-loop $\beta$-function vanishes \cite{Banks:1981nn}, it is possible to construct ${\cal N}=2$ supersymmetric theories finite in all orders \cite{Howe:1983wj}. In particular, ${\cal N}=4$ supersymmetric Yang--Mills theory is finite in all orders \cite{Sohnius:1981sn,Grisaru:1982zh,Howe:1983sr,Mandelstam:1982cb,Brink:1982pd} in agreement with a large number of explicit calculations made in various orders of the perturbation theory \cite{Ferrara:1974pu,Jones:1977zr,Poggio:1977ma,Tarasov:1980au,Tarasov:1980nsq,Avdeev:1981ew,Grisaru:1980nk,Caswell:1980ru,Velizhanin:2010vw}. Some terms breaking ${\cal N}=2$ supersymmetry but preserving the finiteness have been constructed in \cite{Parkes:1982tg,Parkes:1983ib,Parkes:1983nv}.

However, it is important to understand how to calculate quantum corrections in order for the non-renormalization theorems to be valid. For example, the NSVZ $\beta$-relation originally obtained from some general arguments is not satisfied in the $\overline{\mbox{DR}}$ scheme, when a theory is regularized by dimensional reduction \cite{Siegel:1979wq} and divergences are removed by modified minimal subtractions \cite{Bardeen:1978yd}. This was demonstrated by explicit three- and four-loop calculations made in \cite{Jack:1996vg,Jack:1996cn,Jack:1998uj,Harlander:2006xq} (see \cite{Mihaila:2013wma} for review). However, as it turned out, in these approximations the NSVZ equation can be restored with the help of a specially tuned finite renormalization, because its scheme independent consequences \cite{Kataev:2013csa,Kataev:2014gxa} are satisfied. This implies that the NSVZ equation is valid only for special renormalization prescriptions, which are usually called the NSVZ schemes. According to \cite{Goriachuk:2018cac,Goriachuk_Conference,Goriachuk:2020wyn,Korneev:2021zdz}, these schemes constitute an infinite set and are related by finite renormalizations which satisfy a special constraint. A simple prescription giving some NSVZ schemes was obtained in the case of using the higher covariant derivative regularizaton \cite{Slavnov:1971aw,Slavnov:1972sq,Slavnov:1977zf} in the supersymmetric version \cite{Krivoshchekov:1978xg,West:1985jx}. According to \cite{Kataev:2013eta,Stepanyantz:2020uke}, the NSVZ equation is valid in all orders in the HD+MSL scheme.\footnote{In the Abelian case one more all-loop NSVZ prescription is the on-shell scheme \cite{Kataev:2019olb}.} In the HD+MSL scheme a theory is regularized by higher covariant derivatives and divergences are removed by minimal subtraction of logarithms when only powers of $\ln\Lambda/\mu$ (where $\Lambda$ is the dimensionful regularization parameter and $\mu$ is a renormalization scale) are present in the renormalization constants. (The proof was based on the all-order perturbative derivation of the NSVZ $\beta$-function made in \cite{Stepanyantz:2011jy} for ${\cal N}=1$ supersymmetric electrodynamics and in \cite{Stepanyantz:2016gtk,Stepanyantz:2019ihw,Stepanyantz:2020uke} for the ${\cal N}=1$ non-Abelian supersymmetric theories. Its various parts have been verified and confirmed by numerous explicit calculations (see, e.g., \cite{Shakhmanov:2017soc,Shakhmanov:2017wji,Kazantsev:2018nbl,Kuzmichev:2019ywn,Stepanyantz:2019lyo,Aleshin:2020gec,Kuzmichev:2021yjo,Kuzmichev:2021lqa,Aleshin:2022zln}), some of them being made in such orders of the perturbation theory where the scheme dependence becomes essential.)

For ${\cal N}=2$ supersymmetric gauge theories in the $\overline{\mbox{DR}}$ scheme the anomalous dimensions of the chiral matter superfields vanish at least up to the three-loop approximation \cite{Jack:1996qq}. The two- \cite{Howe:1984xq} and three-loop \cite{Jack:1996vg} contributions to the $\beta$-function are also equal to 0. The vanishing of the four-loop contribution to the $\beta$-function of ${\cal N}=2$ supersymmetric gauge theories formulated in terms of ${\cal N}=1$ superfields in the $\overline{\mbox{DR}}$ scheme was an essential ingredient of the calculation made in \cite{Jack:1996cn} where the four-loop $\beta$-function was found for general ${\cal N}=1$ theories except for one undetermined parameter. This in particular implies that the $\overline{\mbox{DR}}$ scheme is NSVZ for ${\cal N}=2$ supersymmetric gauge theories, at least, in the lowest orders.\footnote{Due to the mathematical inconsistency \cite{Siegel:1980qs} dimensional reduction can break supersymmetry in very higher orders \cite{Avdeev:1981vf,Avdeev:1982np,Avdeev:1982xy}.} However, it is known that the finiteness in the $\overline{\mbox{DR}}$ scheme does not in general ensure the finiteness for an arbitrary renormalization prescription. For instance, one-loop finite ${\cal N}=1$ supersymmetric theories in the $\overline{\mbox{DR}}$ scheme are finite in the two-loop approximation \cite{Parkes:1984dh}, but are not two-loop finite for a general ${\cal N}=1$ supersymmetric renormalization prescription \cite{Stepanyantz:2021dus}. Moreover, there are $\overline{\mbox{DR}}$ calculations for ${\cal N}=2$ supersymmetric Yang--Mills theory in the component formulation which reveal the three-loop divergences. Originally these divergences were found in \cite{Avdeev:1982np}. They also remained after the result was corrected in \cite{Velizhanin:2008rw}. Note that for the ${\cal N}=4$ supersymmetric Yang--Mills theory a similar calculation \cite{Velizhanin:2010vw} demonstrated the absence of divergences up to the four-loop approximation. The reason why the three-loop divergences found in \cite{Avdeev:1982np,Velizhanin:2008rw} appear is not now quite clear.

In this paper we consider theories with extended supersymmetry formulated in terms of ${\cal N}=1$ superfields. This implies that one supersymmetry is manifest and survives even at the quantum level, while the others are hidden and can be broken by quantum corrections. In this case expressions for the two-loop anomalous dimensions and for the three-loop $\beta$-function can be found from the corresponding general result for ${\cal N}=1$ supersymmetric gauge theories obtained in \cite{Kazantsev:2020kfl}. Starting from the expressions obtained in this way we analyse the conditions which should be imposed on the renormalization prescription in order for the ${\cal N}=2$ non-renormalization theorem to be valid. In particular, we will demonstrate that the renormalization prescription should be compatible with a structure of quantum corrections, NSVZ, and ${\cal N}=2$ supersymmetric.

The paper is organized as follows. In Sect. \ref{Section_Theory} we recall how gauge theories with extended supersymmetry can be formulated in the ${\cal N}=1$ superspace. The higher covariant derivative regularization, quantization, and renormalization of these theories in a manifestly ${\cal N}=1$ supersymmetric way are performed in Sect. \ref{Section_Quantization}. The two-loop anomalous dimensions and the three-loop $\beta$-function for theories under consideration are investigated in Sect. \ref{Section_Renormalization}. We start with the calculation of the renormalization group functions (RGFs) defined in terms of the bare couplings in Sect. \ref{Subsection_Bare_RGFs} using the general result obtained earlier for ${\cal N}=1$ supersymmetric gauge theories. Next, in Sect. \ref{Subsection_Renormalized_RGFs} we obtain RGFs defined in terms of the renormalized couplings for a general renormalization prescription compatible with ${\cal N}=1$ supersymmetry. A particular case of the $\overline{\mbox{DR}}$ scheme is considered in Sect. \ref{Subsection_DR}. A class of the NSVZ schemes for theories with extended supersymmetry is described in Sect. \ref{Subsection_NSVZ_Schemes}. In Sect. \ref{Section_NonRenormalization_Theorem} we analyse the conditions imposed on the renormalization prescriptions under which the ${\cal N}=2$ and ${\cal N}=4$ non-renormalization theorems are satisfied in the case of using the ${\cal N}=1$ formulation of these theories. In particular, we demonstrate that they are valid for all NSVZ renormalization prescriptions compatible with ${\cal N}=2$ supersymmetry and a structure of quantum corrections. This is verified in the lowest nontrivial approximation in Sect. \ref{Subsection_Nonrenormalization_Lowest} and proved in all orders in Sect. \ref{Subsection_Nonrenormalization_Exact}. In Sect. \ref{Section_Incompatible} we demonstrate the existence of the NSVZ renormalization prescriptions incompatible with a structure of quantum corrections for which the higher order corrections to the anomalous dimensions and the $\beta$-function of ${\cal N}=2$ supersymmetric gauge theories do not vanish. The results of the paper are briefly summarized in Conclusion.

\section{${\cal N}=2$ supersymmetric gauge theories in ${\cal N}=1$ superspace}
\hspace*{\parindent}\label{Section_Theory}

${\cal N}=2$ supersymmetric theories can be considered as a particular case of ${\cal N}=1$ supersymmetric theories and formulated in the ${\cal N}=1$ superspace. In this formulation one supersymmetry is manifest, while the other is hidden. In terms of ${\cal N}=1$ superfields the action of a renormalizable ${\cal N}=2$ supersymmetric gauge theory with a simple gauge group $G$ in the massless limit is given by the expression

\begin{eqnarray}\label{N=2_Theory}
&& S = \frac{1}{2e_0^2} \mbox{Re}\, \mbox{tr} \int d^4x\, d^2\theta\, W^a W_a + \frac{1}{2e_0^2}\mbox{tr}\int d^4x\, d^4\theta\, \Phi^+ e^{2V} \Phi e^{-2V}\nonumber\\
&&\qquad + \frac{1}{4}\int d^4x\, d^4\theta\Big(\phi^+ e^{2V}\phi + \widetilde\phi^+ e^{-2V^T} \widetilde\phi\Big) + \bigg(\frac{i}{\sqrt{2}}\int d^4x\,d^2\theta\,\widetilde\phi^T \Phi \phi + \mbox{c.c.}\bigg).\qquad
\end{eqnarray}

\noindent
Here $V$ is the (Hermitian) gauge superfield with the strength $W_a$. The chiral superfield $\Phi$ in the adjoint representation of the group $G$ is its ${\cal N}=2$ superpartner. The chiral superfields $\phi$ and $\widetilde\phi$ in the representations $R_0$ and $\bar R_0$, respectively, form ${\cal N}=2$ hypermultiplet. In Eq. (\ref{N=2_Theory}) the bare gauge coupling constant is denoted by $e_0$. Below we will also use the notation $\alpha_0 \equiv e_0^2/4\pi$.

In our notation the generators of the gauge group in the fundamental representation are denoted by $t^A$ and satisfy the conditions

\begin{equation}
\mbox{tr}(t^A t^B) = \frac{1}{2} \delta^{AB};\qquad [t^A,t^B] = i f^{ABC} t^C,
\end{equation}

\noindent
where $f^{ABC}$ are (real) structure constants. The generators of the representation $R_0$ we will denote by $T^A$. The similar conditions for them are written as

\begin{equation}
\mbox{tr}(T^A T^B) = T(R_0) \delta^{AB};\qquad [T^A,T^B] = i f^{ABC} T^C.
\end{equation}

\noindent
Also in what follows we will use the notations

\begin{equation}
f^{ACD} f^{BCD} \equiv C_2 \delta^{AB}; \qquad C(R_0)_i{}^j \equiv (T^A T^A)_i{}^j; \qquad r \equiv \mbox{dim}\, G.
\end{equation}

\noindent
In the first two terms of Eq. (\ref{N=2_Theory}) the superfields $V$ and $\Phi$ are expanded in the generators of the fundamental representation, $V = e_0 V^A t^A$, $\Phi = e_0 \Phi^A t^A$, while in the other terms (which contain the superfields $\phi$ and $\widetilde\phi$) it is necessary to use the generators of the representation $R_0$, $V = e_0 V^A T^A$, $\Phi = e_0 \Phi^A T^A$.

The ${\cal N}=4$ supersymmetric Yang--Mills theory is a particular case of the theory (\ref{N=2_Theory}) which corresponds to the hypermultiplet in the adjoint representation of the gauge group, $R_0 = Adj$. In this case $T(R_0) \to T(Adj) = C_2$ and $C(R_0)_i{}^j \to C(Adj)^{IJ} = C_2 \delta^{IJ}$. (Note that we assume that the gauge group is simple.)

The theory (\ref{N=2_Theory}) is invariant under the gauge transformations parameterized by a Lie algebra valued chiral superfield $A$,

\begin{equation}
\phi \to e^{A} \phi;\qquad \widetilde\phi \to e^{-A^T} \widetilde\phi;\qquad \Phi \to e^{A}\Phi e^{-A};\qquad e^{2V} \to e^{-A^+} e^{2V} e^{-A}.
\end{equation}

\noindent
Certainly, it is also invariant under two supersymmetries. One of them is manifest and remains unbroken at the quantum level if the theory is formulated and quantized in terms of ${\cal N}=1$ superfields. The other supersymmetry is hidden. It can be written in the superfield form \cite{Gates:1983nr}

\begin{eqnarray}
&& \delta e^{2V} = i\eta^* e^{2V} \Phi - i\eta \Phi^+ e^{2V};\qquad
\delta\Phi = - \frac{i}{2} W^a D_a\eta;\nonumber\\
&& \delta\phi = - \frac{1}{4\sqrt{2}} \bar D^2\Big(\eta^* e^{-2V} \widetilde\phi^*\Big);\qquad
\delta\widetilde\phi = \frac{1}{4\sqrt{2}} \bar D^2\Big(\eta^* e^{2V^T} \phi^*\Big),
\end{eqnarray}

\noindent
where the parameter $\eta$ is a chiral superfield which does not depend on $x^\mu$. Although the action (\ref{N=2_Theory}) is invariant under these transformations, at the quantum level the hidden supersymmetry can in general be broken in the case of using ${\cal N}=1$ quantization.

As we have already mentioned, the theory (\ref{N=2_Theory}) is a particular case of a general renormalizable ${\cal N}=1$ supersymmetric theory (with a simple gauge group), which in the massless limit is described by the action

\begin{eqnarray}\label{N=1_Action}
&& S = \frac{1}{2e_0^2} \mbox{Re}\, \mbox{tr} \int d^4x\, d^2\theta\, W^a W_a + \frac{1}{4}\int d^4x\, d^4\theta\, \bm{\phi}^{*\bm{i}} (e^{2V})_{\bm{i}}{}^{\bm{j}} \bm{\phi}_{\bm{j}} \nonumber\\
&& \qquad\qquad\qquad\qquad\qquad\qquad\qquad\qquad
+ \Big(\frac{1}{6}\bm{\lambda}_0^{\bm{ijk}} \int d^4x\, d^2\theta\, \bm{\phi}_{\bm{i}} \bm{\phi}_{\bm{j}} \bm{\phi}_{\bm{k}} +\mbox{c.c.}\Big).\qquad
\end{eqnarray}

\noindent
For ${\cal N}=2$ supersymmetric gauge theories the chiral matter superfields $\bm{\phi}_{\bm{i}} = \left(\Phi^A,\, \phi_i,\, \widetilde\phi^i\right)$ belong to the reducible representation

\begin{equation}\label{Representation_R}
R = Adj + R_0 + \bar R_0.
\end{equation}

\noindent
(In this paper we will denote the indices corresponding to the representation $R$ by bold letters. The tensors with such indices (e.g., the matter superfields $\bm{\phi}_{\bm{i}}$ or the bare Yukawa couplings $\bm{\lambda}_0^{\bm{ijk}}$) and group factors corresponding to the representation $R$ we will also indicate by bold letters.) The Yukawa couplings in ${\cal N}=2$ supersymmetric gauge theories are related to the gauge coupling, because

\begin{equation}
\frac{i}{\sqrt{2}}\int d^4x\,d^2\theta\,\widetilde\phi^T \Phi \phi = \frac{ie_0}{\sqrt{2}} (T^A)_i{}^j \int d^4x\,d^2\theta\, \widetilde\phi^i \Phi^A \phi_j.
\end{equation}

\noindent
Comparing this expression with the Yukawa term in the action (\ref{N=1_Action}) we see that the nontrivial components of $\bm{\lambda}_0^{\bm{ijk}}$ are written as

\begin{equation}\label{Coupling_Relation_Bare}
(\bm{\lambda}_0)_i{}^{jA} = (\bm{\lambda}_0)_i{}^{Aj} = (\bm{\lambda}_0)^j{}_i{}^{A} = (\bm{\lambda}_0)^A{}_i{}^{j} = (\bm{\lambda}_0)^{jA}{}_i = (\bm{\lambda}_0)^{Aj}{}_{i} = \frac{ie_0}{\sqrt{2}} (T^A)_i{}^j.
\end{equation}

\section{The higher covariant derivative regularization, quantization, and renormalization in the ${\cal N}=1$ supespace}
\hspace*{\parindent}\label{Section_Quantization}

In this paper we make the regularization and quantization of ${\cal N}=2$ supersymmetric theories in the ${\cal N}=1$ superspace, see, e.g., \cite{Gates:1983nr,West:1990tg,Buchbinder:1998qv}. In this formalism one supersymmetry remains manifest even at the quantum level, while the second (hidden) one can be broken by quantum corrections. The regularization will be made with the help of the Slavnov higher covariant derivative method \cite{Slavnov:1971aw,Slavnov:1972sq} in the superfield formulation \cite{Krivoshchekov:1978xg,West:1985jx}. Note that this regularization in particular includes the insertion of the Pauli--Villars determinants for removing the residual one-loop divergences \cite{Slavnov:1977zf}. The details of this construction in the supersymmetric case can be found in \cite{Aleshin:2016yvj,Kazantsev:2017fdc}.

The choice of the higher covariant derivative regularization is motivated by the fact that the NSVZ equation in supersymmetric theories is valid in all loops in the HD+MSL scheme \cite{Kataev:2013eta,Stepanyantz:2020uke}, so that the $\beta$-function in a certain loop can easily be obtained starting from the expressions for the anomalous dimensions of chiral matter superfields in the previous loops, see, e.g., \cite{Kazantsev:2020kfl,Haneychuk:2022qvu,Shirokov:2022jyd}. Moreover, there are various versions of this regularization, which differ in the form of the higher derivative terms and the Pauli-Villars masses. Therefore, expressions for various RGFs depend on a certain number of regularization parameters, which is very convenient for investigating the scheme dependence.

For quantization of the theories under consideration we will also use the background field method \cite{DeWitt:1965jb,Abbott:1980hw,Abbott:1981ke} formulated in terms of ${\cal N}=1$ superfields \cite{Gates:1983nr}. It is introduced by making the replacement

\begin{equation}
e^{2V} \to e^{2{\cal F}(V)} e^{2\bm{V}},
\end{equation}

\noindent
where $\bm{V}$ denotes the background gauge superfield, and ${\cal F}(V)$ is a certain nonlinear function of the quantum superfield. This function is needed because the quantum gauge superfield is renormalized in a nonlinear way \cite{Piguet:1981fb,Piguet:1981hh,Tyutin:1983rg}. This nonlinear renormalization can be reduced to the linear renormalization of an infinite number of parameters present in ${\cal F}(V)$. The lowest nonlinear term in this function was found in \cite{Juer:1982fb,Juer:1982mp}. It was explicitly demonstrated \cite{Kazantsev:2018kjx} that the renormalization of its coefficient is needed for the renormalization group equations to be satisfied.

Below we will use the general expression for the two-loop anomalous dimension of the chiral matter superfields for ${\cal N}=1$ supersymmetric theories regularized by higher covariant derivatives obtained in \cite{Kazantsev:2020kfl}. That is why here we will use the same version of the higher covariant derivative regularization as in \cite{Kazantsev:2020kfl}. In this version after adding terms with higher derivatives denoted by $S_\Lambda$ the regularized action

\begin{equation}
S_{\mbox{\scriptsize reg}} \equiv S + S_\Lambda
\end{equation}

\noindent
will contain two regulator functions $R(x)$ and $F(x)$ which appear in the kinetic terms for the gauge and matter superfields, respectively. Note that we will use the same regulator function $F(x)$ for all chiral matter superfields of the theory, i.e., $\Phi^A$, $\phi_i$, and $\widetilde\phi^i$.

Constructing the regularized action we also take into account that Eq. (\ref{Coupling_Relation_Bare}) follows from ${\cal N}=2$ supersymmetry, while hidden supersymmetry can in general be broken by quantum corrections. This implies that for a general ${\cal N}=1$ supersymmetric renormalization prescription Eq. (\ref{Coupling_Relation_Bare}) can also be broken. Therefore, if the theory is quantized in the ${\cal N}=1$ superspace, we expect a possible appearance of quantum corrections in which the Yukawa term does not satisfy Eq. (\ref{Coupling_Relation_Bare}). Thus, the regularized action can be written in the form

\begin{eqnarray}\label{Action_Regularized}
&& S_{\mbox{\scriptsize reg}} = \frac{1}{2e_0^2} \mbox{Re}\, \mbox{tr} \int d^4x\, d^2\theta\, W^a \Big[e^{-2\bm{V}} e^{-2{\cal F}(V)} R\Big(-\frac{\bar\nabla^2\nabla^2}{16\Lambda^2}\Big) e^{2{\cal F}(V)} e^{2\bm{V}}\Big]_{Adj} W_a\nonumber\\
&& + \frac{1}{2e_0^2}\mbox{tr}\int d^4x\, d^4\theta\, \Phi^+ \Big[F\Big(-\frac{\bar\nabla^2\nabla^2}{16\Lambda^2}\Big) e^{2{\cal F}(V)} e^{2\bm{V}}\Big]_{Adj} \Phi
\nonumber\\
&& + \frac{1}{4}\int d^4x\, d^4\theta\bigg(\phi^+  F\Big(-\frac{\bar\nabla^2\nabla^2}{16\Lambda^2}\Big) e^{2{\cal F}(V)} e^{2\bm{V}}\phi + \widetilde\phi^+ F\Big(-\frac{\bar\nabla^2\nabla^2}{16\Lambda^2}\Big) e^{-2{\cal F}(V)^T} e^{-2\bm{V}^T} \widetilde\phi\bigg)\qquad
\nonumber\\
&& + \bigg(\frac{ie_0}{\sqrt{2}} \Big[\left(T^A\right)_i{}^j + \left(\lambda_0^A\right)_i{}^j\Big] \int d^4x\,d^2\theta\,\widetilde\phi^i \Phi^A \phi_j + \mbox{c.c.}\bigg),
\end{eqnarray}

\noindent
where the gauge superfield strength is given by the expression

\begin{equation}
W_a = \frac{1}{8}\bar D^2 \Big[e^{-2\bm{V}} e^{-2{\cal F}(V)} D_a \Big(e^{2{\cal F}(V)} e^{2\bm{V}}\Big)\Big],
\end{equation}

\noindent
and a new bare parameter $(\lambda_0^A)_i{}^j$ (of the order $O(\alpha)$) is needed to absorb quantum corrections which break the hidden supersymmetry. In our notation the covariant derivatives are written as

\begin{equation}
\nabla_a = D_a;\qquad \bar\nabla_{\dot a} = e^{2{\cal F}(V)} e^{2\bm{V}} \bar D_{\dot a} e^{-2\bm{V}} e^{-2{\cal F}(V)},
\end{equation}

\noindent
and the subscript $Adj$ indicates that in the corresponding expression the generators should be taken in the adjoint representation,

\begin{equation}
\big(a_0 + a_1 X + a_2 X^2 + \ldots\big)_{Adj} Y \equiv a_0 Y + a_1 [X,Y] + a_2 [X,[X,Y]] +\ldots
\end{equation}

The gauge fixing procedure is made by adding the gauge fixing term $S_{\mbox{\scriptsize gf}}$ and the corresponding Faddeev--Popov and Nielsen--Kallosh ghosts with the actions $S_{\mbox{\scriptsize FP}}$ and $S_{\mbox{\scriptsize NK}}$, respectively. They are rather standard, so that we do not present the corresponding explicit expressions here. They can be found, e.g., in \cite{Kuzmichev:2021lqa}.

The replacement $S\to S_{\mbox{\scriptsize reg}}$ regularizes divergences beyond the one-loop approximation, and the dimensionful parameter $\Lambda$ plays the role of an ultraviolet cutoff. To remove the residual one-loop divergences, following \cite{Slavnov:1977zf}, we insert into the generating functional the Pauli--Villars determinants. According to \cite{Aleshin:2016yvj,Kazantsev:2017fdc}, for this purpose one can use two Pauli--Villars determinants. One of them,

\begin{equation}
\mbox{Det}^{-1}(PV, M_\varphi) = \int D\varphi_1 D\varphi_2 D\varphi_3\, \exp(iS_\varphi),
\end{equation}

\noindent
cancels divergences coming from the (sub)diagrams with one loop of the gauge superfield and ghosts. The action $S_\varphi$ depends on three chiral Pauli--Villars superfields $\varphi_1$, $\varphi_2$, and $\varphi_3$ in the adjoint representation of the gauge group, which have the mass $M_\varphi$ proportional to the parameter $\Lambda$ in the higher derivative term,

\begin{eqnarray}
&& S_\varphi = \frac{1}{2e_0^2} \mbox{tr}\int d^4x\, d^4\theta\, \Big(\varphi_1 ^+ \Big[ R\Big(-\frac{\bar\nabla^2 \nabla^2}{16\Lambda^2}\Big)e^{2{\cal F}(V)}  e^{2\bm{V}}\Big]_{Adj}\varphi_1 + \varphi_2^+ \Big[ e^{2{\cal F}(V)} e^{2\bm{V}}\Big]_{Adj}\varphi_2\qquad\nonumber\\
&& + \varphi_3^+ \Big[ e^{2{\cal F}(V)} e^{2\bm{V}}\Big]_{Adj}\varphi_3\Big) + \frac{1}{2e_0^2}\Big(\mbox{tr}\int d^4x\, d^2\theta\, M_\varphi (\varphi_1^2 + \varphi_2^2 + \varphi_3^2) +\mbox{c.c.}\Big).\qquad
\end{eqnarray}

One more Pauli--Villars determinant removes divergences coming from a loop of chiral matter superfields. For the considered theory it is reasonable to choose it in the form

\begin{equation}
\mbox{Det}^{-1}(PV,M) = \int D\Phi_{PV} D\phi_{PV} D\widetilde\phi_{PV}\, \exp(iS_{PV}),
\end{equation}

\noindent
where the action $S_{PV}$ includes the massive Pauli--Villars superfields $\Phi_{PV}$, $\phi_{PV}$, and $\widetilde\phi_{PV}$ in the representations $Adj$, $R_0$, and $\bar R_0$, respectively, and is written as

\begin{eqnarray}
&&\hspace*{-5mm} S_{PV} = \frac{1}{4}\int d^4x\, d^4\theta\bigg(\phi_{PV}^+  F\Big(-\frac{\bar\nabla^2\nabla^2}{16\Lambda^2}\Big) e^{2{\cal F}(V)} e^{2\bm{V}}\phi_{PV} + \widetilde\phi_{PV}^+ F\Big(-\frac{\bar\nabla^2\nabla^2}{16\Lambda^2}\Big) e^{-2{\cal F}(V)^T} e^{-2\bm{V}^T}
\nonumber\\
&&\hspace*{-5mm} \times \widetilde\phi_{PV}\bigg) + \frac{1}{2e_0^2}\mbox{tr}\int d^4x\, d^4\theta\, \Phi_{PV}^+ \Big[F\Big(-\frac{\bar\nabla^2\nabla^2}{16\Lambda^2}\Big) e^{2{\cal F}(V)} e^{2\bm{V}}\Big]_{Adj} \Phi_{PV} + \bigg( \frac{M}{2}\int d^4x\,d^2\theta\,\Big[\widetilde\phi_{PV}^T \nonumber\\
&&\hspace*{-5mm} \times \phi_{PV} + \frac{1}{e_0^2} \mbox{tr}\left(\Phi_{PV}^2\right)\Big] + \mbox{c.c.}\bigg).
\end{eqnarray}

\noindent
This implies that all these Pauli--Villars superfields have the mass $M$ (which is also proportional to the parameter $\Lambda$). Note that we always assume that the ratios

\begin{equation}
a \equiv \frac{M}{\Lambda};\qquad a_\varphi\equiv \frac{M_\varphi}{\Lambda}
\end{equation}

\noindent
do not depend on the bare couplings. After inserting the above Pauli--Villars determinants the resulting expression for the generating functional can be written as

\begin{equation}
Z[\mbox{sources}] = \int D\mu\,\mbox{Det}(PV,M)\, \exp\Big\{i\Big(S_{\mbox{\scriptsize reg}} + S_{\mbox{\scriptsize gf}} + S_{\mbox{\scriptsize FP}} + S_{\mbox{\scriptsize NK}} + S_\varphi + S_{\mbox{\scriptsize sources}}\Big)\Big\},
\end{equation}

\noindent
where $D\mu$ denotes the functional integration over all superfields of the theory.

For the theory under consideration ultraviolet divergences can be removed by the renormalization of couplings and superfields. The renormalized couplings will be denoted by $e$ (or $\alpha\equiv e^2/4\pi \equiv Z_\alpha\alpha_0$) and $\lambda$. Note that, as we have already mentioned, if ${\cal N}=2$ theories are quantized in ${\cal N}=1$ superspace, then for a general renormalization prescription the ${\cal N}=2$ relation between the gauge and Yukawa couplings is broken by quantum corrections. Due to the absence of divergent quantum corrections to the superpotential \cite{Grisaru:1979wc} the renormalization of $(\lambda^A)_i{}^j$ is related to the renormalization of chiral matter superfields and of the gauge coupling constant by the equation

\begin{equation}
\frac{d}{d\ln\Lambda}\Big[e_0\, \big(Z_\phi^{1/2}\big)_j{}^k\, \big(Z_\phi^{1/2}\big)_l{}^i\, (Z_\alpha Z_\Phi)^{1/2} \Big(\hspace*{-1mm}\left(T^A\right)_i{}^j + \left(\lambda_0^A\right)_i{}^j \Big)\Big] = 0.
\end{equation}

\noindent
In our notation the renormalization constants for the chiral matter superfields are defined as\footnote{Defining the renormalization constant $Z_\Phi$ we follow the notation of Ref. \cite{Buchbinder:2014wra}. In the case of using this definition the corresponding one-loop anomalous dimension vanishes.}

\begin{equation}\label{Renormalization_Constants}
\Phi^A = \big(Z_\alpha Z_\Phi\big)^{1/2} (\Phi_R)^A;\qquad \phi_i = \big(Z_\phi^{1/2}\big)_i{}^j (\phi_{R})_j;\qquad \widetilde\phi^i = \big(Z_\phi^{1/2}\big)_j{}^i (\widetilde\phi_R)^j,
\end{equation}

\noindent
where the subscripts $R$ denote the renormalized superfields. (Note that we consider a theory with a simple gauge group, so that all $\Phi^A$ are renormalized with the same renormalization constant $(Z_\alpha Z_\Phi)^{1/2}$.)

It is convenient to describe ultraviolet divergences with the help of RGFs. According to \cite{Kataev:2013eta}, it is important to distinguish between RGFs defined in terms of the bare couplings and the ones defined in terms of the renormalized couplings,

\begin{eqnarray}\label{RGFs_Definitions}
&& \beta(\alpha_0,\lambda_0) = \frac{d\alpha_0}{d\ln\Lambda}\bigg|_{\alpha,\lambda = \mbox{\scriptsize const}};\qquad\qquad\qquad\ \,
\widetilde\beta(\alpha,\lambda) = \frac{d\alpha}{d\ln\mu}\bigg|_{\alpha_0,\lambda_0 = \mbox{\scriptsize const}}; \nonumber\\
&& (\gamma_\phi)_i{}^j(\alpha_0,\lambda_0) = - \frac{d(\ln Z_\phi)_i{}^j}{d\ln\Lambda}\bigg|_{\alpha,\lambda = \mbox{\scriptsize const}};\qquad
(\widetilde\gamma_\phi)_i{}^j(\alpha,\lambda) = \frac{d(\ln Z_\phi)_i{}^j}{d\ln\mu}\bigg|_{\alpha_0,\lambda_0 = \mbox{\scriptsize const}};\qquad\nonumber\\
&& \gamma_\Phi(\alpha_0,\lambda_0) = - \frac{d\ln Z_\Phi}{d\ln\Lambda}\bigg|_{\alpha,\lambda = \mbox{\scriptsize const}};\qquad\qquad\quad
\widetilde\gamma_\Phi(\alpha,\lambda) = \frac{d\ln Z_\Phi}{d\ln\mu}\bigg|_{\alpha_0,\lambda_0 = \mbox{\scriptsize const}}.\qquad
\end{eqnarray}

\noindent
RGFs defined in terms of the bare couplings are presented in the left column, and RGFs (standardly) defined in terms of the renormalized couplings are presented in the right column. The former ones depend on a regularization, but are independent of a renormalization prescription for a fixed regularization. The latter ones depend on both regularization and renormalization prescription starting from the two-loop approximation for the anomalous dimensions and from the three-loop approximation for the $\beta$-function. RGFs defined in terms of the bare couplings for ${\cal N}=2$ supersymmetric theories are obtained in the case $\lambda_0=0$, and RGFs defined in terms of the renormalized couplings should be calculated at $\lambda=0$. (Nevertheless, the renormalization of the Yukawa couplings should be taken into account.)

Certainly, there is a class of subtraction schemes in which ${\cal N}=2$ supersymmetry survives at the quantum level, so that the Yukawa couplings remain related to the gauge coupling constant and, therefore,

\begin{equation}\label{N=2_Supersymmetric_Scheme}
\frac{d}{d\ln\Lambda}\Big[e_0\, \big(Z_\phi^{1/2}\big)_j{}^k\, \big(Z_\phi^{1/2}\big)_l{}^i\, (Z_\alpha Z_\Phi)^{1/2} \left(T^A\right)_i{}^j \Big] = 0.
\end{equation}

\noindent
Evidently, in this case there is no need to introduce the couplings $\left(\lambda_0^A\right)_i{}^j$, and the anomalous dimensions of the matter superfields are related by the equation

\begin{equation}\label{Superpotential_NonRenormalization}
\gamma_\Phi(\alpha_0) \left(T^A\right)_i{}^j + 2\big(\gamma_\phi\big)_i{}^k(\alpha_0) \left(T^A\right)_k{}^j = 0.
\end{equation}

According to \cite{Stepanyantz:2016gtk,Stepanyantz:2019ihw,Stepanyantz:2020uke}, see also \cite{Stepanyantz:2019lfm}, in the case of using the higher covariant derivative regularization RGFs of ${\cal N}=1$ supersymmetric gauge theories defined in terms of the {\it bare} couplings satisfy the NSVZ equation

\begin{equation}\label{N=1_NSVZ}
\frac{\beta(\alpha_0,\bm{\lambda}_0)}{\alpha_0^2} = - \frac{\left(3 C_2 - \bm{T}(R) + \bm{C}(R)_{\bm{i}}{}^{\bm{j}} \bm{\gamma}_{\bm{j}}{}^{\bm{i}}(\alpha_0,\bm{\lambda}_0)/r\right)}{2\pi(1-\alpha C_2/2\pi)}
\end{equation}

\noindent
for an arbitrary renormalization prescription supplementing this regularization. Here we assume that the chiral matter superfields belong to the representation $R$, for which the generators of the gauge group are denoted by $(\bm{T}^A)_{\bm{i}}{}^{\bm{j}}$. From these generators we construct the group Casimirs

\begin{equation}\label{Casimir_Definitions}
\bm{C}(R)_{\bm{i}}{}^{\bm{j}} \equiv (\bm{T}^A \bm{T}^A)_{\bm{i}}{}^{\bm{j}}; \qquad \mbox{tr}(\bm{T}^A \bm{T}^B) \equiv \bm{T}(R)\delta^{AB}.
\end{equation}

\noindent
For the particular case of ${\cal N}=2$ supersymmetric theories the representation $R$ is given by the direct sum (\ref{Representation_R}) and the Casimirs take the form

\begin{equation}\label{Casimirs}
\bm{T}(R) = C_2 + 2 T(R_0);\qquad \bm{C}(R)_{\bm{i}}{}^{\bm{j}}
= \left(
\begin{array}{ccc}
C_2 \delta^{IJ} & 0 & 0\\
0 & C(R_0)_i{}^j & 0\\
0 & 0 & C(R_0)_j{}^i
\end{array}
\right).
\end{equation}

\noindent
Taking into account that

\begin{equation}\label{Gamma_Phi_Original}
\frac{d\ln(Z_\alpha Z_\Phi)}{d\ln\Lambda}\bigg|_{\alpha,\lambda = \mbox{\scriptsize const}} = - \frac{\beta(\alpha_0,\lambda_0)}{\alpha_0} - \gamma_\Phi(\alpha_0,\lambda_0)
\end{equation}

\noindent
the anomalous dimension matrix can be written as

\begin{equation}\label{Anomalous_Dimension_Matrix}
\bm{\gamma}_{\bm{j}}{}^{\bm{i}} \equiv - \frac{d\ln \bm{Z}_{\bm{j}}{}^{\bm{i}}}{d\ln\Lambda}
= \left(
\begin{array}{ccc}
\left(\beta/\alpha_0 + \gamma_\Phi\right) \delta^{IJ} & 0 & 0\\
0 & \left(\gamma_\phi\right)_j{}^i & 0\\
0 & 0 & \left(\gamma_\phi\right)_i{}^j
\end{array}
\right).
\end{equation}

Substituting the expressions (\ref{Casimirs}) and (\ref{Anomalous_Dimension_Matrix}) into Eq. (\ref{N=1_NSVZ}) after some algebraic transformations we obtain that for the theory under consideration the NSVZ equation takes the form

\begin{equation}\label{N=2_NSVZ}
\frac{\beta(\alpha_0,\lambda_0)}{\alpha_0^2} = - \frac{1}{2\pi}\Big(2C_2 - 2T(R_0) + C_2\gamma_\Phi(\alpha_0,\lambda_0) + 2 C(R_0)_i{}^j (\gamma_\phi)_j{}^i(\alpha_0,\lambda_0)/r\Big).
\end{equation}

\noindent
Note that here we keep the dependence on $\lambda_0$, because (for theories regularized by higher covariant derivatives) this equation is valid for an arbitrary ${\cal N}=1$ supersymmetric theory with the chiral matter superfields in the representation $Adj+R_0+\bar R_0$, and ${\cal N}=2$ supersymmetry is not needed for its derivation.

If we consider a subclass of renormalization prescriptions which satisfy Eq. (\ref{N=2_Supersymmetric_Scheme}) (or, in other words, do not break the relation between the gauge and Yukawa couplings), then it is possible to set $\lambda_0=0$ and express $\gamma_\Phi(\alpha_0)$ in terms of the anomalous dimension of the hypermultiplet by multiplying Eq. (\ref{Superpotential_NonRenormalization}) by $(T^A)_j{}^i$,

\begin{equation}\label{Gamma_Phi_Relation}
\gamma_\Phi(\alpha_0) =  - \frac{2\,\mbox{\small tr}\left(\gamma_\phi(\alpha_0) C(R_0)\right)}{\mbox{\small tr}\, C(R_0)} = - \frac{2\,\mbox{\small tr}\left(\gamma_\phi(\alpha_0) C(R_0)\right)}{r T(R_0)}.
\end{equation}

\noindent
Substituting this expression into Eq. (\ref{N=2_NSVZ}) we obtain the exact $\beta$-function defined in terms of the bare coupling constant \cite{Buchbinder:2014wra},

\begin{equation}
\frac{\beta(\alpha_0)}{\alpha_0^2} = -\frac{1}{\pi}\Big(C_2 - T(R_0)\Big)\Big(1 + \frac{1}{2}\gamma_\Phi(\alpha_0)\Big).
\end{equation}

\noindent
Due to the presence of $\gamma_\Phi(\alpha_0)$ in this equation quantum correction can appear in higher orders. However, if the quantization is made in a manifestly ${\cal N}=2$ supersymmetric way in ${\cal N}=2$ harmonic superspace \cite{Galperin:1984av,Galperin:2001seg,Buchbinder:2001wy}, then this anomalous dimension vanishes \cite{Buchbinder:1997ib}. Then in the case of using the higher covariant derivative regularization formulated in the harmonic superspace \cite{Buchbinder:2015eva} we obtain the ${\cal N}=2$ non-renormalization theorem \cite{Grisaru:1982zh, Howe:1983sr}, according to which the $\beta$-function vanishes beyond the one-loop approximation.

In this paper we will calculate the two-loop anomalous dimensions for the chiral matter superfields and the three-loop $\beta$-function for ${\cal N}=2$ and ${\cal N}=4$ supersymmetric theories formulated in the ${\cal N}=1$ superspace. In particular, we will see that for all NSVZ renormalization prescription compatible with ${\cal N}=2$ supersymmetry and with the structure of quantum corrections these contributions to RGFs vanish beyond the one-loop approximation, so that the ${\cal N}=2$ non-renormalization theorem is valid in the considered order.

\section{The two-loop renormalization of superfields and the three-loop renormalization of the gauge coupling constant}
\label{Section_Renormalization}

\subsection{RGFs defined in terms of the bare couplings}
\hspace*{\parindent}\label{Subsection_Bare_RGFs}

For an arbitrary renormalizable ${\cal N}=1$ supersymmetric gauge theory with a simple gauge group regularized by higher covariant derivatives the general expression for the two-loop anomalous dimension of the matter superfields defined in terms of the bare couplings was obtained in \cite{Kazantsev:2020kfl}. In the notation adopted in this paper it is written as\footnote{The generalization of this expression to the case of theories with multiple gauge couplings can be found in \cite{Haneychuk:2022qvu}.}

\begin{eqnarray}\label{Gamma_Bare}
&&\hspace*{-9mm} \bm{\gamma}_{\bm{i}}{}^{\bm{j}}(\alpha_0,\bm{\lambda}_0) = - \frac{\alpha_0}{\pi}\bm{C}(R)_{\bm{i}}{}^{\bm{j}} + \frac{1}{4\pi^2}\bm{\lambda}^*_{0\bm{imn}} \bm{\lambda}_0^{\bm{jmn}}  + \frac{\alpha_0^2}{2\pi^2} \left[\bm{C}(R)^2\right]_{\bm{i}}{}^{\bm{j}} -\frac{1}{16\pi^4}\bm{\lambda}^*_{0\bm{iac}}\bm{\lambda}_0^{\bm{jab}} \bm{\lambda}^*_{0\bm{bde}} \bm{\lambda}_0^{\bm{cde}}\nonumber\\
&&\hspace*{-3mm} -\frac{3\alpha_0^2}{2\pi^2}\, C_2 \bm{C}(R)_{\bm{i}}{}^{\bm{j}}\Big(\ln a_{\varphi}+1+\frac{A}{2}\Big) +\frac{\alpha_0^2}{2\pi^2}\, \bm{T}(R) \bm{C}(R)_{\bm{i}}{}^{\bm{j}}\Big(\ln a + 1 + \frac{A}{2}\Big) - \frac{\alpha_0}{8\pi^3} \bm{\lambda}^*_{0\bm{lmn}} \bm{\lambda}^{\bm{jmn}}_0 \nonumber\\
&&\hspace*{-3mm} \times \bm{C}(R)_{\bm{i}}{}^{\bm{l}} (1-B+A) + \frac{\alpha_0}{4\pi^3}\bm{\lambda}^*_{0\bm{imn}}\bm{\lambda}_0^{\bm{jml}} \bm{C}(R)_{\bm{l}}{}^{\bm{n}}(1-A+B)
+ O\Big(\alpha_0^3,\alpha_0^2\bm{\lambda}_0^2,\alpha_0\bm{\lambda}_0^4,\bm{\lambda}_0^6\Big).
\end{eqnarray}

\noindent
Here values of the parameters $A$, $B$, $a$, and $a_\varphi$ depend on a particular version of the higher covariant derivative regularization. Namely, the parameters $A$ and $B$ are determined  by the higher derivative regulators $R(x)$ and $F(x)$ present in Eq. (\ref{Action_Regularized}),

\begin{equation}
A \equiv \int\limits_0^\infty dx\, \ln x \frac{d}{dx}\Big(\frac{1}{R(x)}\Big);\qquad B \equiv \int\limits_0^\infty dx\, \ln x \frac{d}{dx}\Big(\frac{1}{F^2(x)}\Big).
\end{equation}

\noindent
The parameters $a \equiv M/\Lambda$ and $a_\varphi \equiv M_\varphi/\Lambda$ are the ratios of the Pauli--Villars masses to the dimensionful parameter of the regularized theory.

Substituting the Yukawa couplings (\ref{Coupling_Relation_Bare}) and the Casimirs (\ref{Casimirs}) into Eq. (\ref{Gamma_Bare}) and taking into account Eq. (\ref{Gamma_Phi_Original}) we obtain 

\begin{eqnarray}\label{Gamma_Phi_Bare_Original}
&&\hspace*{-7mm} \frac{\beta(\alpha_0,\lambda_0=0)}{\alpha_0} + \gamma_\Phi(\alpha_0, \lambda_0 = 0) = - \frac{\alpha_0}{\pi} \Big(C_2 - T(R_0)\Big) + \frac{\alpha_0^2}{\pi^2 r} \mbox{tr}\left[C(R_0)^2\right] (B-A)
\nonumber\\
&&\hspace*{-3mm} + \frac{\alpha_0^2}{2\pi^2} (C_2)^2 \Big(-3\ln a_\varphi + \ln a -1 -A\Big) + \frac{\alpha_0^2}{2\pi^2} C_2 T(R_0) \Big(2\ln a + 1 + B\Big) + O(\alpha_0^3);\\
\label{Gamma_Hyper_Bare}
&&\hspace*{-7mm} (\gamma_\phi)_i{}^j(\alpha_0, \lambda_0 = 0) = \frac{\alpha_0^2}{\pi^2} \left[C(R_0)^2\right]_i{}^j (B-A) + \frac{\alpha_0^2}{2\pi^2} T(R_0) C(R_0)_i{}^j \Big(2\ln a + 1+ A\Big) \nonumber\\
&&\hspace*{-3mm} + \frac{\alpha_0^2}{2\pi^2} C_2\, C(R_0)_i{}^j \Big(-3\ln a_\varphi + \ln a -1 -2A + B\Big) + O(\alpha_0^3).\vphantom{\frac{1}{2}}
\end{eqnarray}

\noindent
Note that in the one-loop approximation both anomalous dimensions vanish. However, in the two-loop approximation this is not true, because the ${\cal N}=1$ regularization and quantization can break the relations following from ${\cal N}=2$ supersymmetry. We see that the two-loop contributions nontrivially depend on the regularization parameters $A$, $B$, $a$, and $a_\varphi$. (Certainly, as we already mentioned above, RGFs defined in terms of the bare couplings are independent of the parameters which determine a subtraction scheme for a fixed regularization.)

The three-loop $\beta$-function defined in terms of the bare couplings can be found using the NSVZ equation (\ref{N=2_NSVZ}). First, from this equation we see that the (scheme independent) two-loop contribution to the $\beta$-function vanishes, because in the one-loop approximation $\gamma_\Phi$ and $(\gamma_\phi)_i{}^j$ are equal to 0. Certainly, this agrees with the well-known non-renormalization theorem \cite{Grisaru:1982zh, Howe:1983sr,Buchbinder:1997ib}. Substituting this result into Eq. (\ref{Gamma_Phi_Bare_Original}) we obtain the two-loop expression for the anomalous dimension $\gamma_\Phi$ (defined in terms of the bare couplings),

\begin{eqnarray}\label{Gamma_Phi_Bare}
&& \gamma_\Phi(\alpha_0, \lambda_0 = 0) = \frac{\alpha_0^2}{\pi^2 r} \mbox{tr}\left[C(R_0)^2\right] (B-A) + \frac{\alpha_0^2}{2\pi^2} (C_2)^2 \Big(-3\ln a_\varphi + \ln a -1 -A\Big) \qquad\nonumber\\
&& \quad + \frac{\alpha_0^2}{2\pi^2} C_2 T(R_0) \Big(2\ln a + 1 + B\Big) + O(\alpha_0^3).
\end{eqnarray}

\noindent
In the case of using the higher covariant derivative regularization Eq. (\ref{N=2_NSVZ}) is valid for RGFs defined in terms of the bare couplings in all orders, and, in particular, relates the three-loop $\beta$-function to the two-loop anomalous dimensions. Substituting the expressions (\ref{Gamma_Hyper_Bare}) and (\ref{Gamma_Phi_Bare}) into the right hand side of Eq. (\ref{N=2_NSVZ}) we obtain

\begin{eqnarray}\label{Beta_Bare}
&& \frac{\beta(\alpha_0,\lambda_0=0)}{\alpha_0^2} = - \frac{1}{\pi}\Big(C_2 - T(R_0)\Big) + \frac{\alpha_0^2}{4\pi^3} (C_2)^3 \Big(3\ln a_\varphi - \ln a + 1 + A\Big)
-\frac{\alpha_0^2}{4\pi^3 r} (C_2)^2 \qquad\nonumber\\
&&\quad \times \mbox{tr}\, C(R_0) \Big(2\ln a + 1 +B\Big) + \frac{\alpha_0^2}{2\pi^3 r} C_2\, \mbox{tr}\left[C(R_0)^2\right] \Big(3\ln a _\varphi - \ln a + 1 +3A -2B\Big)
\nonumber\\
&&\quad + \frac{\alpha_0^2}{\pi^3 r} \mbox{tr}\left[C(R_0)^3\right] (A-B)
- \frac{\alpha_0^2}{2\pi^3 r^2} \mbox{tr}\, C(R_0)\, \mbox{tr}\left[C(R_0)^2\right] \Big(2\ln a + 1 + A\Big) + O(\alpha_0^3).
\end{eqnarray}

\noindent
Equivalently, this expression can be derived from the general equation presented in \cite{Kazantsev:2020kfl}. According to Eq. (\ref{Beta_Bare}), the (regularization dependent) three-loop contribution to the $\beta$-function is not equal to 0, again, because ${\cal N}=1$ regularization and quantization can break the relations following from ${\cal N}=2$ supersymmetry. From Eq. (\ref{Gamma_Hyper_Bare}) and (\ref{Gamma_Phi_Bare}) we also see that Eq. (\ref{Superpotential_NonRenormalization}) is not in general valid. Therefore, for a general ${\cal N}=1$ supersymmetric renormalization prescription it is really necessary to introduce the parameter $\lambda_0$ into the action, see Eq. (\ref{Action_Regularized}).

\subsection{RGFs defined in terms of the renormalized couplings}
\hspace*{\parindent}\label{Subsection_Renormalized_RGFs}

Next, we proceed to calculating RGFs defined in terms of the renormalized couplings. For this purpose we first integrate the renormalization group equations in the left column of Eq. (\ref{RGFs_Definitions}). The solutions contain some finite constants which fix a renormalization prescription in the considered order of the perturbation theory. For example, the relation between the bare and renormalized gauge coupling constants can be presented in the form

\begin{eqnarray}\label{Alpha_Scheme}
&& \frac{1}{\alpha_0} = \frac{1}{\alpha} + \frac{C_2}{\pi}\Big(\ln\frac{\Lambda}{\mu} + b_{11}\Big) - \frac{T(R_0)}{\pi}\Big(\ln\frac{\Lambda}{\mu} + b_{12}\Big) \nonumber\\
&&\qquad + \frac{\alpha}{\pi^2} (C_2)^2 b_{21} - \frac{\alpha}{2\pi^2 r} C_2\, \mbox{tr}\, C(R_0) b_{22} - \frac{\alpha}{\pi^2 r} \mbox{tr}\left[C(R_0)^2\right] b_{23} + O(\alpha^2,\alpha\lambda), \qquad
\end{eqnarray}

\noindent
where $b_i$ are the finite constants. Note that we included into this expression all products of group factors which can appear in the considered approximation. This implies that we deal with subtraction schemes compatible with the structure of quantum corrections \cite{Jack:1996vg,Jack:2016tpp}.\footnote{All renormalization prescriptions considered in this paper are also compatible with ${\cal N}=1$ supersymmetry because the quantization and renormalization are made in terms of ${\cal N}=1$ superfields.} Similarly, the renormalization constants for the matter superfields in the lowest approximation contain the finite constants $j_i$ and $g_i$ and can be written as

\begin{eqnarray}\label{Z_Phi_Scheme}
&& Z_\Phi = 1 + \frac{\alpha}{\pi} C_2 j_{11} - \frac{\alpha}{\pi} T(R_0) j_{12} + O(\alpha^2,\alpha\lambda);\qquad\\
\label{Hypermultiplet_Scheme}
&& \big(Z_\phi\big)_i{}^j = \delta_i^j + \frac{\alpha}{\pi} C(R_0)_i{}^j g_{1} + O(\alpha^2,\alpha\lambda).\vphantom{\frac{1}{2}}
\end{eqnarray}

\noindent
Also finite constants $l_i$ can appear in the renormalization of the coupling $\lambda_0$ present in Eq. (\ref{Action_Regularized}). In the lowest approximation the expression $\lambda_0-\lambda$ (where $\lambda$ is the corresponding renormalized coupling) should be finite. Taking into account all possible structures that can appear in calculating quantum corrections in ${\cal N}=1$ theories it is possible to present the relation between the bare and renormalized Yukawa couplings in the lowest order in the form

\begin{eqnarray}\label{Lambda_Scheme}
&& \left(\lambda_0^A\right)_i{}^j = \left(\lambda^A\right)_i{}^j - \frac{\alpha C_2}{2\pi} \left(T^A\right)_i{}^j\,  l_{11} + \frac{\alpha T(R_0)}{2\pi} \left(T^A\right)_i{}^j\, l_{12} \nonumber\\
&&\qquad\qquad\qquad\qquad\qquad\qquad\qquad  + \frac{\alpha}{2\pi} C\big(R_0\big)_i{}^k \left(T^A\right)_k{}^j\, l_{13} + O\big(\alpha^2,\alpha\lambda\big).\qquad
\end{eqnarray}

\noindent
Note that usually the renormalization of the Yukawa couplings in supersymmetric theories is made according to the prescription

\begin{equation}\label{Yukawa_Standard_Renormalization}
\bm{\lambda}^{\bm{ijk}} = \bm{\lambda}_0^{\bm{mnp}} (\sqrt{\bm{Z}})_{\bm{m}}{}^{\bm{i}} (\sqrt{\bm{Z}})_{\bm{n}}{}^{\bm{j}} (\sqrt{\bm{Z}})_{\bm{p}}{}^{\bm{k}},
\end{equation}

\noindent
where ${\bm{Z}}_{\bm{i}}{}^{\bm{j}}$ are the renormalization constants for the chiral matter superfields, $\bm{\phi}_{\bm{i}} = (\sqrt{\bm{Z}})_{\bm{i}}{}^{\bm{j}} (\bm{\phi}_R)_{\bm{j}}$. For the theory under consideration this subtraction scheme corresponds to the finite constants satisfying the constraints

\begin{equation}\label{Yukawa_Standard_Finite_Constants}
l_{11} = j_{11}; \qquad l_{12} = j_{12}; \qquad l_{13} = -2g_1.
\end{equation}

\noindent
However, below we will use a more general renormalization prescription in which the coefficients $l_i$ are arbitrary.

Substituting the renormalization constants for the matter superfields and the relation between the bare and renormalized coupling constants into the equations presented in the right column of Eq. (\ref{RGFs_Definitions}) we obtain RGFs defined in terms of the renormalized couplings. Note that for ${\cal N}=2$ supersymmetric gauge theories they should be calculated at $\lambda = 0$, but the renormalization of the Yukawa couplings should nevertheless be taken into account, because we consider general ${\cal N}=1$ supersymmetric renormalization prescriptions, which, in particular, include the ones breaking ${\cal N}=2$ supersymmetry. The resulting expressions for RGFs are written as

\begin{eqnarray}\label{N=2_Gamma_Phi}
&&\hspace*{-7mm} \widetilde\gamma_\Phi(\alpha,\lambda=0) = \frac{\alpha^2}{\pi^2 r} \mbox{tr}\left[C(R_0)^2\right] \Big(B-A+l_{13}\Big) + \frac{\alpha^2}{2\pi^2} (C_2)^2 \Big(-3\ln a_\varphi + \ln a -1 -A - 2 j_{11}\Big) \nonumber\\
&&\hspace*{-3mm} + \frac{\alpha^2}{2\pi^2} C_2\, T(R_0) \Big(2\ln a + 1 + B - 2 l_{11} + 2 j_{11} + 2 j_{12}\Big) + \frac{\alpha^2}{\pi^2} T(R_0)^2 \Big(l_{12}-j_{12}\Big) + O(\alpha^3);\\
&&\hspace*{-3mm} \nonumber\\
\label{N=2_Gamma_Hypermultiplet}
&& \hspace*{-7mm} \big(\widetilde\gamma_\phi\big)_i{}^j(\alpha,\lambda=0) = \frac{\alpha^2}{\pi^2}\left[C(R_0)^2\right]_i{}^j \Big(B-A+l_{13}\Big)
+ \frac{\alpha^2}{2\pi^2}\, T(R_0) C\big(R_0\big)_i{}^j \Big(2\ln a + 1 + A + 2 g_{1}\nonumber\\
&&\hspace*{-3mm} + 2 l_{12}\Big)
+ \frac{\alpha^2}{2\pi^2} C_2\, C\big(R_0\big)_i{}^j \Big(-3\ln a_\varphi + \ln a -1 - 2A + B - 2g_{1} - 2 l_{11}\Big) + O(\alpha^3);\\
&&\hspace*{-3mm} \nonumber\\
\label{N=2_Beta}
&&\hspace*{-7mm} \frac{\widetilde\beta(\alpha,\lambda=0)}{\alpha^2} = - \frac{1}{\pi}\Big(C_2 - T(R_0)\Big) + \frac{\alpha^2}{4\pi^3} (C_2)^3 \Big(3\ln a_\varphi - \ln a + 1+A - 4b_{21} \Big)
-\frac{\alpha^2}{4\pi^3 r} (C_2)^2\nonumber\\
&&\hspace*{-3mm} \times \mbox{tr}\, C(R_0)\, \Big(2\ln a + 1+ B - 2b_{22} - 2l_{11}-4b_{21}\Big)
+ \frac{\alpha^2}{2\pi^3 r} C_2\, \mbox{tr}\left[C(R_0)^2\right] \Big(3\ln a_\varphi - \ln a \nonumber\\
&&\hspace*{-3mm} + 3A -2B +1 + 2b_{23} + 2l_{11} - l_{13}\Big) - \frac{\alpha^2}{2\pi^3 r^2} C_2 \left[\mbox{tr}\,C(R_0)\right]^2 \Big(b_{22} + l_{12}\Big)
+ \frac{\alpha^2}{\pi^3 r} \mbox{tr}\left[C(R_0)^3\right] \nonumber\\
&&\hspace*{-3mm} \times \Big(A - B - l_{13}\Big) - \frac{\alpha^2}{2\pi^3 r^2} \mbox{tr}\, C(R_0)\, \mbox{tr}\left[C(R_0)^2\right] \Big(2\ln a + A + 1 + 2 b_{23}
+ 2l_{12}\Big) + O(\alpha^3) \vphantom{\frac{1}{2}}
\end{eqnarray}

\noindent
and depend on both regularization parameters and finite constants which determine a renormalization prescription.

For the ${\cal N}=4$ supersymmetric Yang--Mills theory $R_0 = Adj$, so that $T(R_0) = C_2$ and $C(R_0)_i{}^j \to C_2 \delta^{IJ}$. Therefore, in this case RGFs defined in terms of the renormalized couplings take the form

\begin{eqnarray}\label{N=4_Gamma_Phi}
&&\hspace*{-7mm} \widetilde\gamma_\Phi(\alpha,\lambda=0) =  \frac{\alpha^2}{2\pi^2} (C_2)^2 \Big(3B-3A -3\ln \frac{a_\varphi}{a} - 2 l_{11} +2l_{12} +2l_{13}\Big) + O(\alpha^3);\\
&&\vphantom{1}\nonumber\\
\label{N=4_Gamma_Hypermultiplet}
&& \hspace*{-7mm} \big(\widetilde\gamma_\phi\big)^{AB}(\alpha,\lambda=0) = \delta^{AB}\cdot \frac{\alpha^2}{2\pi^2} (C_2)^2 \Big(3B-3A -3\ln \frac{a_\varphi}{a} - 2 l_{11} + 2 l_{12} + 2l_{13}\Big) + O(\alpha^3);\qquad\\
&&\vphantom{1}\nonumber\\
\label{N=4_Beta}
&&\hspace*{-7mm} \frac{\widetilde\beta(\alpha,\lambda=0)}{\alpha^2} = -\frac{3\alpha^2}{4\pi^3} (C_2)^3 \Big(3B -3A -3\ln \frac{a_\varphi}{a} -2l_{11} +2l_{12} + 2l_{13}\Big)
+ O(\alpha^3).
\end{eqnarray}

\noindent
From these equations we see that both the three-loop $\beta$-function and the two-loop anomalous dimensions do not in general vanish. This seems to contradict the well-known fact that the ${\cal N}=4$ supersymmetric Yang--Mills theory is finite in all loops \cite{Sohnius:1981sn,Grisaru:1982zh,Howe:1983sr,Mandelstam:1982cb,Brink:1982pd}. However, we actually considered the theory with manifest ${\cal N}=1$ supersymmetry and {\it admitted such renormalizations that spoil extended supersymmetry}. If we restrict ourselves to such renormalization prescriptions that do not break extended supersymmetry, then the theory will be finite. We will discuss this in detail below in Sect. \ref{Section_NonRenormalization_Theorem}. Here we will only note that for an arbitrary renormalization prescription the expressions (\ref{N=4_Gamma_Phi}) --- (\ref{N=4_Beta}) satisfy the equations

\begin{eqnarray}\label{N=4_Gamma_Relation_Lowest}
&& \big(\widetilde\gamma_\phi\big)^{AB}(\alpha,\lambda=0) = \delta^{AB} \widetilde\gamma_\Phi(\alpha,\lambda=0) + O(\alpha^3);\vphantom{\frac{1}{2}}\qquad\\
\label{N=4_Beta_Relation_Lowest}
&& \frac{\widetilde\beta(\alpha,\lambda=0)}{\alpha^2} = - \frac{3}{2\pi} C_2 \widetilde\gamma_\Phi(\alpha,\lambda=0) + O(\alpha^3).
\end{eqnarray}

\subsection{RGFs in the $\overline{\mbox{DR}}$ scheme}
\hspace*{\parindent}\label{Subsection_DR}

Let us compare the results for RGFs obtained above with the corresponding expressions in the $\overline{\mbox{DR}}$ scheme found in \cite{Jack:1996vg,Jack:1996cn,Jack:1998uj}. The expressions for finite constants corresponding to this renormalization prescription were found in \cite{Kazantsev:2020kfl},  where the relation between the bare and renormalized coupling constant and the renormalization constant for the chiral matter superfields were written in the form

\begin{eqnarray}\label{Two_Loop_Alpha_General}
&&\hspace*{-5mm} \frac{1}{\alpha} - \frac{1}{\alpha_0} = -\frac{3}{2\pi}C_2\Big(\ln \frac{\Lambda}{\mu} + \bm{b}_{11}\Big) + \frac{1}{2\pi} \bm{T}(R) \Big(\ln \frac{\Lambda}{\mu} + \bm{b}_{12}\Big) - \frac{3\alpha}{4\pi^2} (C_2)^2 \Big(\ln\frac{\Lambda}{\mu} + \bm{b}_{21} \Big) \qquad\nonumber\\
&&\hspace*{-5mm} \quad + \frac{\alpha}{4\pi^2 r}  C_2 \mbox{\bf tr}\, \bm{C}(R) \Big(\ln \frac{\Lambda}{\mu} + \bm{b}_{22}\Big) + \frac{\alpha}{2\pi^2 r} \mbox{\bf tr}\left[\bm{C}(R)^2\right] \Big(\ln \frac{\Lambda}{\mu} + \bm{b}_{23}\Big) - \frac{1}{8\pi^3 r} \bm{C}(R)_{\bm{j}}{}^{\bm{i}}\nonumber\\
&&\hspace*{-5mm} \quad \times \bm{\lambda}^*_{\bm{imn}} \bm{\lambda^{jmn}} \Big(\ln \frac{\Lambda}{\mu}  + \bm{b}_{24}\Big) + O(\alpha^2,\alpha \bm{\lambda}^2,\bm{\lambda}^4); \vphantom{\frac{1}{2}}\\
\label{One_Loop_Z_General}
&&\hspace*{-5mm} \bm{Z}_{\bm{i}}{}^{\bm{j}}(\alpha,\bm{\lambda}) = \delta_{\bm{i}}{}^{\bm{j}} + \frac{\alpha}{\pi} \bm{C}(R)_{\bm{i}}{}^{\bm{j}} \Big(\ln\frac{\Lambda}{\mu}+\bm{g}_{11}\Big) - \frac{1}{4\pi^2} \bm{\lambda}^*_{\bm{imn}}\bm{\lambda}^{\bm{jmn}} \Big(\ln\frac{\Lambda}{\mu} + \bm{g}_{12} \Big) + O(\alpha^2,\alpha\bm{\lambda}^2,\bm{\lambda}^4).\nonumber\\
\end{eqnarray}

\noindent
For the considered ${\cal N}=2$ supersymmetric gauge theories considered in this paper the Yukawa couplings are given by Eq. (\ref{Coupling_Relation_Bare}). Substituting them into Eqs. (\ref{Two_Loop_Alpha_General}) and (\ref{One_Loop_Z_General}) and comparing the result with Eqs. (\ref{Alpha_Scheme}), (\ref{Z_Phi_Scheme}), and (\ref{Hypermultiplet_Scheme}) we establish the correspondence between the notations of Ref. \cite{Kazantsev:2020kfl} and of this paper,

\begin{eqnarray}\label{B_Correspondence}
&&\hspace*{-5mm} g_1 = \bm{g}_{11} - \bm{g}_{12};\qquad j_{11} = \bm{g}_{11} -\frac{3}{2} \bm{b}_{11} + \frac{1}{2} \bm{b}_{12};\qquad\ \ \, j_{12} = \bm{g}_{12} - \bm{b}_{12};\qquad\, b_{11} = \frac{3}{2} \bm{b}_{11} - \frac{1}{2} \bm{b}_{12};\quad  \nonumber\\
&&\hspace*{-5mm} b_{12} = \bm{b}_{12}; \qquad\qquad\ b_{21} = \frac{3}{4} \bm{b}_{21} - \frac{1}{4} \bm{b}_{22} - \frac{1}{2} \bm{b}_{23};\qquad\,
b_{22} = \bm{b}_{22} - \bm{b}_{24};\qquad\,
b_{23} = \bm{b}_{23} - \bm{b}_{24}.\qquad
\end{eqnarray}

According to \cite{Kazantsev:2020kfl}, for a general ${\cal N}=1$ supersymmetric gauge theory regularized by higher covariant derivatives RGFs (defined in terms of the renormalized couplings) coincide with the ones in the $\overline{\mbox{DR}}$ scheme if the finite constants are given by the expressions

\begin{eqnarray}\label{B_DR}
&& \bm{b}_{11} = \ln a_\varphi;\qquad\qquad \bm{b}_{12} = \ln a;\qquad\qquad \bm{g}_{11} = -\frac{1}{2} - \frac{A}{2};\qquad \bm{g}_{12} = - \frac{1}{2} - \frac{B}{2};\qquad \nonumber\\
&& \bm{b}_{21} = \ln a_\varphi + \frac{1}{4};\qquad \bm{b}_{22} = \ln a + \frac{1}{4};\qquad\, \bm{b}_{23} = -\frac{1}{4} - \frac{A}{2};\qquad \bm{b}_{24} = -\frac{1}{4} - \frac{B}{2}.\qquad
\end{eqnarray}

\noindent
Also in the $\overline{\mbox{DR}}$ scheme the renormalization of the Yukawa couplings is made according to Eq. (\ref{Yukawa_Standard_Renormalization}), so that the corresponding finite constants are given by Eq. (\ref{Yukawa_Standard_Finite_Constants}). Thus, using Eqs.  (\ref{Yukawa_Standard_Finite_Constants}), (\ref{B_Correspondence}), and (\ref{B_DR}) we obtain the values of the finite constants corresponding to the
$\overline{\mbox{DR}}$ scheme,

\begin{eqnarray}\label{DR_Finite_Constants}
&& b_{11} = \frac{3}{2}\ln a_\varphi -\frac{1}{2}\ln a;\qquad\qquad\qquad\ \, b_{12} = \ln a; \qquad\qquad\qquad\quad\ \, g_1 = \frac{1}{2}(B-A);\qquad \nonumber\\
&& b_{21} = \frac{3}{4}\ln a_\varphi - \frac{1}{4}\ln a + \frac{1}{4}+\frac{A}{4};\qquad\
b_{22} = \ln a + \frac{1}{2} + \frac{B}{2};\qquad\quad\ b_{23} = \frac{1}{2}(B-A);\nonumber\\
&& l_{11} = -\frac{3}{2}\ln a_\varphi + \frac{1}{2}\ln a -\frac{1}{2} - \frac{A}{2};\quad\ \ \, l_{12} = -\ln a - \frac{1}{2} - \frac{B}{2};\qquad\ \, l_{13} = A-B;\nonumber\\
&& j_{11} = -\frac{3}{2}\ln a_\varphi + \frac{1}{2}\ln a -\frac{1}{2} - \frac{A}{2};\quad\ \ \, j_{12} = -\ln a - \frac{1}{2} - \frac{B}{2}.
\end{eqnarray}

Substituting these values of the finite constants into Eqs. (\ref{N=2_Gamma_Phi}) --- (\ref{N=2_Beta}) we obtain RGFs in the $\overline{\mbox{DR}}$ scheme,

\begin{eqnarray}
&& \widetilde\gamma_\Phi(\alpha,\lambda=0) = O(\alpha^3);\vphantom{\frac{1}{2}}\qquad
\big(\widetilde\gamma_\phi\big)_i{}^j(\alpha,\lambda=0) = O(\alpha^3);\qquad\\
&& \frac{\widetilde\beta(\alpha,\lambda=0)}{\alpha^2} = - \frac{1}{\pi}\Big(C_2 - T(R_0)\Big) + O(\alpha^3). \vphantom{\frac{1}{2}}
\end{eqnarray}

\noindent
This implies that the ${\cal N}=2$ (and, therefore, ${\cal N}=4$) non-renormalization theorems are satisfied in the $\overline{\mbox{DR}}$ scheme at least in the considered apprximation.

\subsection{NSVZ schemes for ${\cal N}=2$ supersymmetric theories}
\hspace*{\parindent}\label{Subsection_NSVZ_Schemes}

Note that RGFs (\ref{N=2_Gamma_Phi}) --- (\ref{N=2_Beta}) satisfy the NSVZ equation

\begin{equation}\label{N=2_NSVZ_Renormalized}
\frac{\widetilde\beta(\alpha,\lambda=0)}{\alpha^2} = - \frac{1}{2\pi}\Big(2C_2 - 2T(R_0) + C_2\widetilde\gamma_\Phi(\alpha,\lambda=0) + 2 C(R_0)_i{}^j (\widetilde\gamma_\phi)_j{}^i(\alpha,\lambda=0)/r\Big)
\end{equation}

\noindent
only if the finite constants fixing a renormalization prescription satisfy the equations

\begin{equation}\label{NSVZ_Finite_Constants}
2 b_{21} + j_{11} = 0;\qquad b_{22} + j_{12} = 0;\qquad b_{23} - g_1 = 0,
\end{equation}

\noindent
which specify the class of NSVZ schemes. From Eq. (\ref{DR_Finite_Constants}) it is easy to see that the $\overline{\mbox{DR}}$ scheme is NSVZ for ${\cal N}=2$ supersymmetric theories in agreement with \cite{Jack:1996vg,Jack:1996cn}.

Eq. (\ref{NSVZ_Finite_Constants}) agrees with the general statement \cite{Goriachuk:2018cac,Goriachuk_Conference,Goriachuk:2020wyn,Korneev:2021zdz} that (for RGFs defined in terms of the renormalized couplings) various NSVZ schemes are related by finite renormalizations $\alpha' = \alpha'(\alpha)$, $(\bm{Z}')_{\bm{i}}{}^{\bm{j}} = (\bm{z})_{\bm{i}}{}^{\bm{k}} (\bm{Z})_{\bm{k}}{}^{\bm{j}}$ which satisfy the constraint

\begin{equation}\label{NSVZ_Class_Constraint}
\frac{1}{\alpha'} - \frac{1}{\alpha} + \frac{C_2}{2\pi} \ln \frac{\alpha'}{\alpha} - \frac{1}{2\pi r} \bm{C}(R)_{\bm{i}}{}^{\bm{j}} (\ln \bm{z})_{\bm{j}}{}^{\bm{i}} = \bm{B},
\end{equation}

\noindent
where $\bm{B}$ is a finite constant. For ${\cal N}=2$ supersymmetric gauge theories considered in this paper Eq. (\ref{NSVZ_Class_Constraint}) takes the form

\begin{equation}
\frac{1}{\alpha'} - \frac{1}{\alpha} -\frac{1}{\pi r} C(R_0)_i{}^j (\ln z_\phi)_j{}^i - \frac{1}{2\pi} C_2 \ln z_\Phi = \bm{B},
\end{equation}

\noindent
where $(z_\phi)_i{}^j$ and $\alpha' z_\Phi/\alpha$ describe the finite renormalizations of the hypermultiplet superfields and of the chiral superfields $\Phi^A$, respectively.\footnote{According to Eq. (\ref{Renormalization_Constants}), in our notation the renormalization constants for the superfields $\Phi^A$ are $(Z_\alpha Z_\Phi)^{1/2} = (\alpha Z_\Phi/\alpha_0)^{1/2}$.} In the case of using the higher covariant derivative regularization some NSVZ schemes are given by the HD+MSL prescription \cite{Kataev:2013eta}, when divergences are removed by minimal subtractions of logarithms.\footnote{Minimal subtractions of logarithms can supplement various versions of the higher covariant derivative regularization, so that there is a certain set of the HD+MSL schemes, each of them being NSVZ.} If the values without primes correspond to this scheme, then the scheme defined by Eqs. (\ref{Alpha_Scheme}) --- (\ref{Lambda_Scheme}) is obtained after the finite renormalization

\begin{eqnarray}
&& \frac{1}{\alpha'} = \frac{1}{\alpha} - \frac{C_2}{\pi} b_{11} + \frac{T(R_0)}{\pi} b_{12} - \frac{\alpha}{\pi^2} (C_2)^2 b_{21}\nonumber\\
&&\qquad\qquad\qquad\qquad\qquad + \frac{\alpha C_2}{2\pi^2 r}\, \mbox{tr}\, C(R_0) b_{22} + \frac{\alpha}{\pi^2 r} \mbox{tr}\left[C(R_0)^2\right] b_{23} + O(\alpha^2,\alpha\lambda); \qquad\nonumber\\
&& \left(\lambda'{}^A\right)_i{}^j = \left(\lambda^A\right)_i{}^j + \frac{\alpha C_2}{2\pi} \left(T^A\right)_i{}^j\,  l_{11} - \frac{\alpha T(R_0)}{2\pi} \left(T^A\right)_i{}^j\, l_{12} \nonumber\\
&&\qquad\qquad\qquad\qquad\qquad\qquad\qquad  - \frac{\alpha}{2\pi} C\big(R_0\big)_i{}^k \left(T^A\right)_k{}^j\, l_{13} + O\big(\alpha^2,\alpha\lambda\big);\qquad\nonumber\\
&& z_\Phi = 1 + \frac{\alpha}{\pi} C_2 j_{11} - \frac{\alpha}{\pi} T(R_0) j_{12} + O(\alpha^2,\alpha\lambda);\qquad\nonumber\\
&& \big(z_\phi\big)_i{}^j = \delta_i^j + \frac{\alpha}{\pi} C(R_0)_i{}^j g_{1} + O(\alpha^2,\alpha\lambda). \vphantom{\frac{1}{2}}
\end{eqnarray}

\noindent
Substituting these expressions into Eq. (\ref{NSVZ_Class_Constraint}) we obtain the constraints (\ref{NSVZ_Finite_Constants}) together with the equation

\begin{equation}
\bm{B} = - \frac{C_2}{\pi} b_{11} + \frac{T(R_0)}{\pi} b_{12},
\end{equation}

\noindent
which determines the constant $\bm{B}$.

\section{The non-renormalization theorems}
\label{Section_NonRenormalization_Theorem}

\subsection{The lowest nontrivial approximation}
\hspace*{\parindent}\label{Subsection_Nonrenormalization_Lowest}

The results for RGFs presented above seem to disagree with the well-known non-renormalization theorems for theories with extended supersymmetry. Namely, in ${\cal N}=2$ supersymmetric gauge theories the $\beta$-function should contain only the one-loop contribution \cite{Grisaru:1982zh, Howe:1983sr,Buchbinder:1997ib}, and all anomalous dimensions should vanish \cite{Grisaru:1982zh, Howe:1983sr}. The ${\cal N}=4$ supersymmetric Yang--Mills theory should be finite in all orders \cite{Sohnius:1981sn,Grisaru:1982zh,Howe:1983sr,Mandelstam:1982cb,Brink:1982pd}. As we have already mentioned, the contradiction appears because we admit such renormalization prescriptions that break extended supersymmetry. Therefore, in general, we cannot expect that the above non-renormalization theorems will be valid. However, there are special classes of subtraction schemes which are compatible with extended supersymmetry. In this section we construct such renormalization prescriptions and verify for them the validity of the non-renormalization theorems.

In theories with ${\cal N}=2$ supersymmetry the Yukawa couplings are related to the gauge coupling constant. Written in terms of the bare couplings this relation is given by Eq. (\ref{Coupling_Relation_Bare}). If we use a renormalization prescription compatible with ${\cal N}=2$ supersymmetry, then a similar equation should also be valid for the renormalized couplings. In particular, this implies that it is possible to choose such a subtraction scheme that

\begin{equation}\label{N=2_Charge_Renormalization}
e_0 \left(T^A\right)_l{}^k = e\, \big(Z_\phi^{-1/2}\big)_j{}^k\, \big(Z_\phi^{-1/2}\big)_l{}^i\, (Z_\alpha Z_\Phi)^{-1/2} \left(T^A\right)_i{}^j.
\end{equation}

\noindent
Differentiating Eq. (\ref{N=2_Charge_Renormalization}) with respect to $\ln\mu$ at fixed values of the bare couplings we obtain the equation

\begin{equation}\label{RGFs_Relation_Renormalized}
\widetilde\gamma_\Phi(\alpha) \left(T^A\right)_i{}^j = - 2\big(\widetilde\gamma_\phi\big)_i{}^k(\alpha) \left(T^A\right)_k{}^j.
\end{equation}

Substituting RGFs (\ref{N=2_Gamma_Phi}) --- (\ref{N=2_Beta}) into this relation and equating coefficients at various group factors we obtain the constraints on the finite constants fixing a renormalization prescription  compatible with ${\cal N}=2$ supersymmetry in the considered approximation,\footnote{Strictly speaking, Eq. (\ref{N=2_Charge_Renormalization}) is more restrictive and, in particular, in the lowest order also gives the equation $g_1=(B-A)/2$. However, below we will investigate only consequences of the weaker condition (\ref{RGFs_Relation_Renormalized}).}

\begin{eqnarray}\label{N=2_Scheme}
&& l_{11}= \frac{B}{2} - \frac{A}{2} +j_{11} - g_{1};\qquad\quad\ \ l_{12} = j_{12};\qquad\quad\ \  l_{13} = A-B; \nonumber\\
&& j_{11} = -\frac{3}{2}\ln a_\varphi +\frac{1}{2}\ln a - \frac{1}{2} - \frac{A}{2};\qquad
j_{12} = -\ln a - \frac{1}{2} - \frac{A}{2} - g_{1}.\qquad
\end{eqnarray}

\noindent
(Note that the values of finite constants (\ref{DR_Finite_Constants}) corresponding to the $\overline{\mbox{DR}}$-scheme satisfy these equations.) It is easy to see that the constraints (\ref{N=2_Scheme}) lead to the vanishing of the two-loop contributions to the anomalous dimensions of all chiral matter superfields, so that

\begin{equation}\label{N=2_Gamma_Trivial}
\widetilde\gamma_\Phi(\alpha) = O(\alpha^3);\qquad \left(\widetilde\gamma_\phi\right)_i{}^j(\alpha) = O(\alpha^3).
\end{equation}

However, the three-loop contribution to the $\beta$-function does not in general vanish,

\begin{eqnarray}\label{N=2_Beta_NON_NSVZ}
&& \frac{\widetilde\beta(\alpha)}{\alpha^2} = - \frac{1}{\pi}\Big(C_2 - T(R_0)\Big) \bigg[1 - \frac{\alpha^2}{\pi^2 r} \mbox{tr}\left(C(R_0)^2\right)\Big(b_{23}- g_{1}\Big)
\nonumber\\
&&\qquad\quad\ \ + \frac{\alpha^2}{2\pi^2} (C_2)^2 \Big(2b_{21} + j_{11}\Big) - \frac{\alpha^2}{2\pi^2} C_2 T(R_0) \Big(b_{22} + j_{12}\Big)\bigg] + O(\alpha^3).\qquad
\end{eqnarray}

\noindent
The vanishing three-loop contribution is obtained only in the NSVZ schemes, which satisfy Eq. (\ref{NSVZ_Finite_Constants}) and, in particular, include the $\overline{\mbox{DR}}$ scheme.\footnote{Certainly, we discuss only  theories with extended supersymmetry. For ${\cal N}=1$ supersymmetric theories in the $\overline{\mbox{DR}}$ scheme the NSVZ equation is not in general satisfied \cite{Jack:1996vg,Jack:1996cn,Jack:1998uj}.} Thus, in the considered approximation the ${\cal N}=2$ non-renormalization theorem is valid in all ${\cal N}=2$ supersymmetric NSVZ schemes (compatible with the structure of quantum corrections),

\begin{equation}
\frac{\widetilde\beta(\alpha)}{\alpha^2} = - \frac{1}{\pi}\Big(C_2 - T(R_0)\Big) + O(\alpha^3).
\end{equation}

For the ${\cal N}=4$ supersymmetric Yang-Mills theory the hypermultiplet superfields belong to the adjoint representation, and the theory is invariant under the manifest $SO(3)$ symmetry which rotates chiral matter superfields. Taking into account that, according to Eq. (\ref{Renormalization_Constants}), the renormalization constant for the superfields $\Phi^A$ contains the factor $(Z_\alpha)^{1/2}$ we conclude that for renormalization prescriptions compatible with the $SO(3)$ symmetry the anomalous dimensions should be related by the equation

\begin{equation}\label{N=4_Gamma_Constraint}
(\widetilde\gamma_\phi)^{AB}(\alpha) = \delta^{AB}\Big(\widetilde\gamma_\Phi(\alpha) + \frac{\widetilde\beta(\alpha)}{\alpha}\Big).
\end{equation}

\noindent
This equation agrees with Eq. (\ref{N=4_Gamma_Relation_Lowest}), because the (scheme independent) two-loop $\beta$-function for the model under consideration is equal to 0. Note that the validity of Eq. (\ref{N=4_Gamma_Relation_Lowest}) is not so trivial, because, in general, we choose different renormalization constants for the superfield $\Phi$ and for the hypermultiplet superfields. Nevertheless, the finite constants present in $Z_\Phi$ and $(Z_\phi)_i{}^j$ did not enter the expressions for the two-loop anomalous dimensions of the ${\cal N}=4$ theory (given by Eqs. (\ref{N=4_Gamma_Phi}) and (\ref{N=4_Gamma_Hypermultiplet})), so that the $SO(3)$ symmetry really leads to the coincidence of all anomalous dimensions in this approximation.

For ${\cal N}=4$ renormalization prescriptions the ${\cal N}=4$ relation between the gauge and Yukawa couplings

\begin{equation}
\lambda^{iA,jB,kC} = -\frac{e}{\sqrt{2}} f^{ABC} \varepsilon^{ijk}
\end{equation}

\noindent
(where the $SO(3)$ indices $i,j,k$ take the values from 1 to 3 and numerate three chiral matter superfields in the adjoint representation) should be valid for the renormalized values. Therefore, from Eqs. (\ref{RGFs_Relation_Renormalized}) and (\ref{N=4_Gamma_Constraint}) we obtain the relation\footnote{Note that in our notation for ${\cal N}=4$ supersymmetric Yang--Mills theory and renormalization prescriptions compatible with the $SO(3)$ symmetry $\Phi_i^A = (Z_\alpha Z_\Phi)^{1/2} (\Phi_R)^A_i$.}

\begin{equation}\label{N=4_Beta_Constraint}
\qquad\widetilde\beta(\alpha) = -\frac{3\alpha}{2}\widetilde\gamma_\Phi(\alpha).
\end{equation}

In the lowest nontrivial approximation ($O(\alpha^3)$) the left hand side of Eq. (\ref{N=4_Beta_Constraint}) vanishes, and we obtain the relation between finite constants which should be satisfied for the renormalization prescriptions compatible with ${\cal N}=4$ supersymmetry,

\begin{equation}
3B-3A -3\ln \frac{a_\varphi}{a} - 2 l_{11} +2l_{12} +2l_{13} = 0.
\end{equation}

\noindent
Then from Eqs. (\ref{N=4_Gamma_Phi}) --- (\ref{N=4_Beta}) we see that all RGFs vanish in the considered approximation,

\begin{equation}
\widetilde\gamma_\Phi = O(\alpha^3);\qquad (\widetilde\gamma_\phi)^{AB} = O(\alpha^3);\qquad \frac{\widetilde\beta(\alpha)}{\alpha^2} = O(\alpha^3),
\end{equation}

\noindent
certainly, in agreement with the ${\cal N}=4$ non-renormalization theorem. Note that, according to Eqs. (\ref{N=2_Gamma_Trivial}) and (\ref{N=2_Beta_NON_NSVZ}) in the considered approximation this theorem will be valid even for an arbitrary ${\cal N}=2$ renormalization prescription.

\subsection{The all-loop results}
\hspace*{\parindent}\label{Subsection_Nonrenormalization_Exact}

Evidently, for ${\cal N}=2$ supersymmetric gauge theories the anomalous dimension $(\widetilde\gamma_\phi)_i{}^j$ should include group structures with two indices constructed from the generators of the representation $R_0$ and the gauge group structure constants. Therefore, the expression $(\widetilde\gamma_\phi)_i{}^k  (T^A)_k{}^j$ can contain only terms proportional to $(C_2)^x \left[C(R_0)^y\right]_i{}^k (T^A)_k{}^j$, where $x\ge 0$, $y\ge 1$. If we admit only renormalization prescriptions compatible with the structure of quantum corrections and with ${\cal N}=2$ supersymmetry, then we should equate the coefficients at different group structures in Eq. (\ref{RGFs_Relation_Renormalized}). However, its left side is proportional to $(T^A)_i{}^j$, while the right hand side cannot contain this structure. Therefore, both sides of this equation should vanish, so that in this case

\begin{equation}
\widetilde\gamma_\Phi(\alpha) = 0;\qquad \big(\widetilde\gamma_\phi\big)_i{}^j(\alpha) = 0.
\end{equation}

\noindent
Then from the NSVZ equation written for renormalization prescriptions which do not break Eq. (\ref{RGFs_Relation_Renormalized}),

\begin{equation}\label{N=2_Relaxed_NSVZ}
\widetilde\beta(\alpha) = -\frac{\alpha^2}{\pi}\Big(C_2 - T(R_0)\Big)\Big(1 + \frac{1}{2}\widetilde\gamma_\Phi(\alpha)\Big),
\end{equation}

\noindent
we obtain that only the one-loop contribution to the $\beta$-function does not vanish,

\begin{equation}
\widetilde\beta(\alpha) = -\frac{\alpha^2}{\pi}\Big(C_2 - T(R_0)\Big).
\end{equation}

Now, it is possible to formulate the conditions under which the ${\cal N}=2$ non-renormalization theorems are valid for ${\cal N}=2$ theories formulated in ${\cal N}=1$ superspace. Namely, the anomalous dimensions of chiral matter superfields vanish if

1. The renormalization prescription does not break the ${\cal N}=2$ relation between the gauge and Yukawa couplings;

2. The renormalization prescription is compatible with the structure of quantum corrections.

3. Moreover, all contributions to the $\beta$-function beyond the one-loop approximation vanish if the conditions 1 and 2 are satisfied and the renormalization prescription is NSVZ.

For ${\cal N}=4$ supersymmetric Yang--Mills theory from Eqs.  (\ref{N=2_NSVZ_Renormalized}) and (\ref{N=4_Gamma_Constraint}) we conclude that for an arbitrary scheme compatible with $SO(3)$ symmetry the NSVZ equation takes the form

\begin{equation}\label{N=4_NSVZ}
\frac{\widetilde\beta(\alpha)}{\alpha^2} = - \frac{3C_2 \widetilde\gamma_\Phi(\alpha)}{2\pi(1+\alpha C_2/\pi)},
\end{equation}

\noindent
where the factor 3 appears because the theory contains three chiral superfields in the adjoint representation of the gauge group. According to \cite{Stepanyantz:2021dus}, for one-loop finite ${\cal N}=1$ supersymmetric theories the NSVZ equation should be satisfied in the first nontrivial approximation {\it for an arbitrary renormalization prescription}. Eq. (\ref{N=4_Beta_Relation_Lowest}) exactly confirms this statement.

If we will consider ${\cal N}=4$ supersymmetric schemes, then from Eq. (\ref{N=4_Beta_Constraint}) (which is valid in this case) and the NSVZ relation (\ref{N=4_NSVZ}) we obtain the all-loop finiteness of the theory under consideration,

\begin{equation}
\widetilde\gamma_{\Phi_i}(\alpha) = 0;\qquad \widetilde\beta(\alpha) = 0,
\end{equation}

\noindent
where all chiral matter superfields of the theory are denoted by $\Phi_i^A$ with $i=1,2,3$. Note that in this case there is even no need to require the validity of the NSVZ equation. Really, according to \cite{Parkes:1985hj,Grisaru:1985tc} (see also \cite{Stepanyantz:2021dus}), for ${\cal N}=1$ supersymmetric theories finite in a certain loop the $\beta$-function vanishes in the next loop. Then, according to Eq. (\ref{N=4_Beta_Constraint}) the anomalous dimension of the chiral matter superfields in this next loop also vanishes. This certainly implies that for ${\cal N}=4$ supersymmetric schemes the ${\cal N}=4$ supersymmetric Yang--Mills theory is finite in all loops.

Certainly, the ${\cal N}=4$ non-renormalization theorem is also valid if the renormalization prescription satisfies the above conditions 1 --- 3.

\section{Renormalization prescriptions incompatible with the structure of quantum corrections}
\hspace*{\parindent}\label{Section_Incompatible}

In the previous section we demonstrated that for any ${\cal N}=2$ supersymmetric renormalization prescription compatible with the structure of quantum corrections the anomalous dimensions of all chiral matter superfields vanish. Also for the ${\cal N}=2$ supersymmetric NSVZ schemes compatible with the structure of quantum corrections all contributions to the $\beta$-function beyond the one-loop approximation are equal to 0. However, according to  \cite{Buchbinder:2014wra}, the NSVZ equation and the relation between the Yukawa and gauge couplings do not ensure that the anomalous dimensions vanish and only the one-loop contribution to the $\beta$-function is nontrivial. In this section we will demonstrate that for the renormalization prescriptions which are incompatible with the structure of quantum corrections the higher order corrections to RGFs can be nontrivial even for the NSVZ schemes.

For simplicity, below we will assume that the hypermultiplet representation $R_0$ is irreducible. Therefore, the hypermultiplet anomalous dimension is proportional to the identity matrix,

\begin{equation}
\big(\widetilde\gamma_\phi\big)_i{}^j = \widetilde\gamma_\phi \delta_i^j.
\end{equation}

\noindent
Similarly, we obtain

\begin{equation}
C(R_0)_i{}^j = C(R_0)\delta_i^j\qquad\mbox{with}\qquad C(R_0) = \frac{r}{d} T(R_0),
\end{equation}

\noindent
where $r = \delta^{AA}$ is the dimension of the gauge group, and $d = \delta_i^i$ is the dimension of the representation $R_0$. Then the result of \cite{Buchbinder:2014wra} for the exact $\beta$-function obtained for the NSVZ renormalization prescriptions which do not break the relation between the gauge and Yukawa couplings is given by Eq. (\ref{N=2_Relaxed_NSVZ}). Now, let us find out if it is really possible to obtain nontrivial higher order corrections to the gauge $\beta$-function for such renormalization prescriptions. Earlier we saw that the anomalous dimension in the right hand side vanishes if a renormalization prescription is compatible with a structure of quantum corrections. However, if this is not so, then the anomalous dimension and the higher order contributions to the $\beta$-function can be nontrivial even for ${\cal N}=2$ supersymmetric NSVZ renormalization prescriptions. How can this occur? Let us assume that we calculate quantum corrections for an ${\cal N}=2$ supersymmetric theory with a fixed gauge group and a fixed hypermultiplet representation (unlike the calculation described above, when the gauge group and hypermultiplet representation were arbitrary). In this case {\it all group factors become numbers}, and we cannot distinguish between, say, $C_2$ and $T(R_0)$. Then, according to Eqs. (\ref{Alpha_Scheme}) --- (\ref{Lambda_Scheme}), the renormalization prescription in the lowest loops is determined by the finite constants

\begin{eqnarray}\label{Erased_Finite_Constants}
&& j_1 \equiv C_2 j_{11} - T(R_0) j_{12};\qquad\quad l_1 \equiv C_2 l_{11} - T(R_0) l_{12} - \frac{r}{d} T(R_0) l_{13};\qquad  \nonumber\\
&& b_1 \equiv C_2 b_{11} - T(R_0) b_{12};\qquad\quad
b_2 \equiv (C_2)^2 b_{21} - \frac{1}{2} C_2 T(R_0) b_{22} - \frac{r}{d}\, T(R_0)^2 b_{23},
\end{eqnarray}

\noindent
and $g_1$. In this section by an explicit calculation we demonstrate that these finite constants can be chosen so that the two-loop contributions to the anomalous dimensions of the chiral superfields and the three-loop contribution to the $\beta$-function do not vanish even for ${\cal N}=2$ supersymmetric NSVZ renormalization prescriptions. Really, for an irreducible hypermultiplet representation RGFs (\ref{N=2_Gamma_Phi}) --- (\ref{N=2_Beta}) can be rewritten in terms of the finite constants (\ref{Erased_Finite_Constants}) as

\begin{eqnarray}\label{N=2_Gamma_Phi_Irreducible}
&&\hspace*{-3mm} \widetilde\gamma_\Phi(\alpha,\lambda=0) =  - \frac{3\alpha^2}{2\pi^2} (C_2)^2 \ln \frac{a_\varphi}{a} + \frac{\alpha^2}{2\pi^2} T(R_0) \Big(C_2 + \frac{2r}{d} T(R_0)\Big)  (B-A) - \frac{\alpha^2}{2\pi^2} \Big(C_2 - T(R_0)\Big) \nonumber\\
&&\hspace*{-3mm} \times C_2 \Big(2\ln a + 1 + A\Big) -\frac{\alpha^2}{\pi^2} T(R_0) \Big(l_1-j_1\Big) - \frac{\alpha^2}{\pi^2} C_2 j_1 + O(\alpha^3);\\
&&\vphantom{1}\nonumber\\
\label{N=2_Gamma_Hypermultiplet_Irreducible}
&&\hspace*{-3mm} \widetilde\gamma_\phi(\alpha,\lambda=0) = - \frac{3\alpha^2 r}{2\pi^2 d} C_2\, T(R_0) \ln \frac{a_\varphi}{a} + \frac{\alpha^2 r}{2\pi^2 d}  T(R_0) \Big(C_2 + \frac{2r}{d} T(R_0)\Big) (B-A) - \frac{\alpha^2 r}{2\pi^2 d} T(R_0)
\nonumber\\
&&\hspace*{-3mm} \times \Big(C_2 - T(R_0)\Big)  \Big(2\ln a +1 + A\Big) - \frac{\alpha^2 r}{\pi^2 d} T(R_0) \Big(C_2 - T(R_0)\Big) g_1  -\frac{\alpha^2 r}{\pi^2 d} T(R_0) l_1 + O(\alpha^3);\\
&& \vphantom{1}\nonumber\\
\label{N=2_Beta_Irreducible}
&&\hspace*{-3mm} \frac{\widetilde\beta(\alpha,\lambda=0)}{\alpha^2} = - \frac{1}{\pi}\Big(C_2 - T(R_0)\Big) + \frac{3\alpha^2}{4\pi^3} C_2 \Big((C_2)^2 + \frac{2r}{d} T(R_0)^2\Big) \ln \frac{a_\varphi}{a}
+ \frac{\alpha^2}{4\pi^3} \Big(C_2-T(R_0)\Big) \nonumber\\
&&\hspace*{-3mm} \times \Big((C_2)^2 + \frac{2r}{d} T(R_0)^2\Big) \Big(2 \ln a + 1 + A\Big)
-\frac{\alpha^2}{4\pi^3} T(R_0) \Big(C_2 + \frac{2r}{d} T(R_0)\Big)^2 (B - A) + \frac{\alpha^2}{2\pi^3} T(R_0) \nonumber\\
&&\hspace*{-3mm} \times \Big(C_2 + \frac{2r}{d}\, T(R_0)\Big) l_1 -\frac{\alpha^2}{\pi^3} \Big(C_2-T(R_0)\Big) b_2
+ O(\alpha^3), \vphantom{\frac{1}{2}}
\end{eqnarray}

\smallskip

\noindent
and the NSVZ conditions (\ref{NSVZ_Finite_Constants}) take the form

\begin{equation}\label{Erased_NSVZ_Finite_Constants}
b_2 = - \frac{1}{2} C_2 j_1 - \frac{r}{d}\, T(R_0)^2 g_1.
\end{equation}

\noindent
Due to Eq. (\ref{RGFs_Relation_Renormalized}) in the case of the irreducible representation $R_0$ for ${\cal N}=2$ supersymmetric renormalization prescriptions RGFs satisfy the relation

\begin{equation}\label{RGFs_Relation_Renormalized_Irreducible}
\widetilde\gamma_\Phi(\alpha) = - 2\widetilde\gamma_\phi(\alpha).
\end{equation}

\noindent
Earlier we saw that if this condition is satisfied separately for all various group structures present in Eqs. (\ref{N=2_Gamma_Phi}) --- (\ref{N=2_Beta}), then the two-loop anomalous dimensions vanish. This implies that the considered ${\cal N}=2$ supersymmetric renormalization scheme is compatible with the structure of quantum corrections. However, now we consider a fixed gauge group and a fixed hypermultiplet representation, so that it is impossible to distinguish between various group structures. Therefore, the condition (\ref{RGFs_Relation_Renormalized_Irreducible}) is not so restrictive and in the considered approximation gives only the relation

\begin{eqnarray}\label{Erased_N=2_Scheme}
&& T(R_0) l_1 = - \frac{d C_2 + 2r T(R_0)}{2(d+2r)} \bigg[ 3 C_2 \ln \frac{a_\varphi}{a} + \Big(C_2 - T(R_0)\Big) \Big(2\ln a + 1 + A\Big) \qquad \nonumber\\
&&\qquad\qquad\qquad\qquad - \Big(1+\frac{2r}{d}\Big) T(R_0) (B-A)\bigg] - \frac{C_2-T(R_0)}{d+2r} \Big(2r T(R_0) g_1 + d\, j_1\Big).\qquad
\end{eqnarray}

Substituting the value of the finite constant $b_2$ from the NSVZ condition (\ref{Erased_NSVZ_Finite_Constants}) and the value of the finite constant $l_1$ from Eq. (\ref{Erased_N=2_Scheme}) into Eqs. (\ref{N=2_Gamma_Phi_Irreducible}) --- (\ref{N=2_Beta_Irreducible}) we obtain the expressions for RGFs in ${\cal N}=2$ supersymmetric schemes (which are in general incompatible with the structure of quantum corections),

\begin{eqnarray}
&& \widetilde\gamma_\Phi(\alpha,\lambda=0) = - 2\widetilde\gamma_\phi(\alpha,\lambda=0)= \frac{\alpha^2 r}{\pi^2 (d+2r)} \Big( C_2 - T(R_0)\Big) \bigg[ -3C_2 \ln\frac{a_\varphi}{a} \nonumber\\
&&\qquad\qquad\qquad\qquad\quad   - \Big( C_2 - T(R_0)\Big)\Big(2\ln a + 1 + A\Big) + 2T(R_0) g_1 - 2 j_1\bigg] + O(\alpha^3);\qquad\\
&&\vphantom{1}\nonumber\\
&& \frac{\widetilde\beta(\alpha,\lambda=0)}{\alpha^2} = - \frac{1}{\pi}\Big(C_2 - T(R_0)\Big)
- \frac{\alpha^2 r}{2\pi^3 (d+2r)} \Big( C_2 - T(R_0)\Big)^2 \bigg[ -3C_2 \ln\frac{a_\varphi}{a} \nonumber\\
&&\qquad\qquad\qquad\qquad\quad   - \Big( C_2 - T(R_0)\Big)\Big(2\ln a + 1 + A\Big) + 2T(R_0) g_1 - 2 j_1\bigg]
+ O(\alpha^3).
\end{eqnarray}

\noindent
We see that they satisfy the relation (\ref{RGFs_Relation_Renormalized_Irreducible}) and the NSVZ equation (\ref{N=2_Relaxed_NSVZ}), but the two-loop contributions to the anomalous dimensions and the three-loop contribution to the $\beta$-function do not in general vanish. This implies that for ${\cal N}=2$ NSVZ supersymmetric schemes {\it which are not compatible with the structure of quantum corrections} the ${\cal N}=2$ non-renormalization theorem can in general be broken. However, for renormalization prescriptions which satisfy the condition

\begin{equation}
j_1 - T(R_0) g_1 + \frac{3}{2} C_2 \ln \frac{a_\varphi}{a} + \frac{1}{2} \Big( C_2 - T(R_0)\Big) \Big(2\ln a + 1 + A\Big) = 0
\end{equation}

\noindent
the two-loop contributions to the anomalous dimensions and the three-loop contribution to the $\beta$-function vanish, and this theorem is valid.

\section{Conclusion}
\hspace*{\parindent}

In this paper we analysed quantum corrections in $D=4$ gauge theories with extended supersymmetry formulated in ${\cal N}=1$ superspace. In this formulation the only supersymmetry is manifest at the quantum level, while the others are hidden and can in general be broken at the quantum level. That is why in general we cannot expect that the non-renormalization theorems following from the extended supersymmetry are satisfied. This was confirmed by the explicit calculations of the two-loop anomalous dimensions and of the three-loop $\beta$-function made for a general renormalization prescription compatible with ${\cal N}=1$ supersymmetry supplementing the higher covariant derivative regularization. The choice of this regularization was motivated by the fact that in this case all NSVZ schemes can easily be constructed. In particular, some of these schemes (which differ in the choice of the regulator functions and the Pauli--Villars masses) are given by the HD+MSL prescription, when divergences are removed by minimal subtractions of logarithms.

We demonstrated that for a general renormalization prescription the two-loop anomalous dimension of the chiral matter superfields and the three-loop contribution to the $\beta$-function do not vanish. Moreover, the equation (\ref{RGFs_Relation_Renormalized}) can also be broken by quantum corrections. As we discussed, this occurs because the ${\cal N}=2$ relation (\ref{Coupling_Relation_Bare}) between the Yukawa couplings and the gauge coupling constant is satisfied only in the case of using the renormalization prescriptions compatible with ${\cal N}=2$ supersymmetry, while in general a subtraction scheme can break it. By other words, the nonrenormalization of superpotential determines the evolution of the Yukawa couplings, but in the case of ${\cal N}=1$ supersymmetric quantization it is possible that their renormalization group behaviour does not coincide with the evolution of the gauge coupling for renormalization prescriptions breaking ${\cal N}=2$ supersymmetry.

The three-loop contribution to the $\beta$-function of ${\cal N}=2$ supersymmetric gauge theories is not in general equal to 0. This contribution vanishes if a renormalization prescription is NSVZ, does not break ${\cal N}=2$ supersymmetry, and is compatible with the structure of quantum corrections. It was demonstrated that under these conditions the anomalous dimensions of chiral superfields vanish in all orders and all contributions to the $\beta$-function beyond the one-loop approximation are equal to 0 in agreement with the well-known non-renormalization theorem \cite{Grisaru:1982zh, Howe:1983sr,Buchbinder:1997ib}. Note that the $\overline{\mbox{DR}}$ scheme satisfies the above conditions at least in the three-loop approximation. In particular, in this approximation it is NSVZ for theories with extended supersymmetry.

However, if a renormalization prescription is incompatible with the structure of quantum corrections, it is possible to construct such ${\cal N}=2$ supersymmetric NSVZ renormalization schemes that RGFs do not vanish in higher orders. An example of such a scheme was explicitly constructed in this paper.

For ${\cal N}=4$ supersymmetric Yang--Mills theory divergences can appear if a renormalization prescription breaks extended supersymmetry exactly as in ${\cal N}=2$ theories. In particular, it was demonstrated that for a general ${\cal N}=1$ substraction scheme the two-loop anomalous dimension and the three-loop $\beta$-function may be different from to 0. However, these contributions to RGFs satisfy the NSVZ equation for any renormalization prescription according to the general theorem \cite{Stepanyantz:2021dus} which states that for one-loop finite theories the NSVZ equation in the first nontrivial order is satisfied for any subtraction scheme compatible with the finiteness in the previous loops. However, for an arbitrary renormalization prescription compatible with ${\cal N}=4$ supersymmetry the ${\cal N}=4$ supersymmetric Yang--Mills theory appears to be finite in agreement with \cite{Sohnius:1981sn,Grisaru:1982zh,Howe:1983sr,Mandelstam:1982cb,Brink:1982pd}.

Thus, we formulated the conditions under which the non-renormalization theorems are valid for theories with extended supersymmetry formulated in ${\cal N}=1$ superspace.

\section*{Acknowledgments}
\hspace*{\parindent}

The work of K.S. was supported by Russian Scientific Foundation, grant No. 21-12-00129.

\end{document}